\documentclass[twocolumn,secnumarabic,amssymb, nobibnotes, nofootinbib,aps,pra]{revtex4-1}

\usepackage[%
  colorlinks=true,
  urlcolor=blue,
  linkcolor=blue,
  citecolor=blue
]{hyperref}

\usepackage{graphicx}
\usepackage{amsmath}
\usepackage{epsfig}
\usepackage{bm}
\usepackage[usenames]{color}
\usepackage{lipsum} % To generate some text for the article via \lipsum

\def\>{{\rangle}}
\def\<{{\langle}}

\def\bs{\boldsymbol }

\begin{document}

\title{Coherent single-spin electron resonance spectroscopy   manifested at an exceptional-point singularity in a doped polyacetylene}

% Author Orchid ID: enter ID or remove command
\newcommand{\orcidauthorA}{0000-0002-1057-2585} % Add \orcidA{} behind the author's name

\author{Yujin \surname{Dunham}}

\affiliation{Department of Physical Science, Osaka Prefecture University, Gakuen-cho
1-1, Sakai 599-8531, Japan}

\author{Kazuki \surname{Kanki}}

\affiliation{Department of Physical Science, Osaka Prefecture University, Gakuen-cho
1-1, Sakai 599-8531, Japan}

\author{Savannah \surname{Garmon}$^*$}
\email{sgarmon@p.s.osakafu-u.ac.jp}
\affiliation{Department of Physical Science, Osaka Prefecture University, Gakuen-cho
1-1, Sakai 599-8531, Japan}

\author{Gonzalo \surname{Ordonez}}

\affiliation{Department of Physics, Butler University, 4600 Sunset Ave., Indianapolis
IN 46208, USA}

\author{Satoshi \surname{Tanaka}$^\dagger$}
\email{stanaka@p.s.osakafu-u.ac.jp}
\affiliation{Department of Physical Science, Osaka Prefecture University, Gakuen-cho
1-1, Sakai 599-8531, Japan}

\small

%############################

\begin{abstract}

Spin-dependent charge transfer decay in an alkali atom doped polyacetylene is studied in terms of the complex spectral analysis, revealing the single-spin Zeeman splitting influenced by the spin-orbit interaction.
Nonhermitian effective Hamiltonian has been derived from the total system hermitian Hamiltonian using Brillouin-Wigner-Feshbach projection method, where the microscopic spin-dependent dissipation effect is correctly incorporated in the energy-dependent self-energy.
Since the present method maintains the dynamical and chiral symmetries of the total system, we discovered two types of exceptional point (EP) singularities in a unified perspective: the EP surface and EP ring are attributed to the dynamical and chiral symmetry breaking, respectively.

We have revealed that the coherent single-spin electron resonance (SSESR) spectrum reflects the complex eigenenergy spectrum of the system.
We have formulated the SSESR spectrum in terms of the nonlinear response function in the Liouville-space pathway approach, where we have constructed the Liouville space basis using the complex eigenstates of the total Hamiltonian.

We have calculated the one- and two-dimensional Fourier transform SSESR (1DFT and 2DFT) spectra reflecting the spin-relaxation dynamics at the donor site.
While the 1DFT SSESR spectrum reflects the complex eigenenergy spectrum, the 2DFT gives detailed information on the quantum coherence in the spin-relaxation dynamics as a cross-correlation between the two frequencies.
We found a giant response of the coherent SSESR around the EP ring singularity due to the vanishing normalization factors at the EP ring and the resonance effect.
We have discovered that the giant response is much larger in magnitudes in the 2DFT spectrum than in the 1DFT spectrum, which promises the 2DFT SSESR   a useful tool to observe the single-spin response in a molecule.

\end{abstract}

\date{\today}
\maketitle

%#################
\section{Introduction}

Conventional idea that dissipation is just a minor perturbation causing population decay or quantum decoherence has  been overturned in recent years.
It has been recognized that the dissipation process has a siginificant impact on the quantum evolution of a state, especially when it is close to an accidental degeneracy point, such as a {\it diabolic point} (DP).
In contrast to ordinary perturbation, which removes the degeneracy in energy,  dissipation gives rise to a peculiar singularity, called {\it exceptional-point singularity} (EP), where the eigenvalues and the eigenstates coalesce \cite{Hatano96PRL,Bender98PRL,Bender1999,BenderPTsymmetry,kato2012short,Mostafazadeh2015}.

Many intriguing phenomena attributed to the EP singularities have been reported, such as asymmetric mode conversion\cite{Yoon2018},  unidirectional invisibility\cite{Lin2011}, dynamical control\cite{Garmon2017a}, higher order phase transition with Fano resonance\cite{Tanaka16PRA},  a large amplification of the Fano absorption spectral component\cite{Fukuta17PRA}, amplification of lasing\cite{Feng2014Science}, enhanced nonlinear response to a perturbation\cite{Peng14NatPhys,Wiersig2014,Chen2017,Hodaei2017,Pick2017,Yu2020}, and amplified spontaneous emission\cite{Lin2016,Pick2017a}.
Since the system energy changes abruptly with a perturbation parameter around the EP singularity, the nonlinear response to an external field can be dramatically  enhanced.

Recently, there have been many efforts to explore various materials exhibiting characteristic behaviors of the EP singularities\cite{BenderPTsymmetry}.
As the number of the inner degrees of freedom of a subsystem increases, such as spin variables, polarizations in electronic and/or photonic systems, the number of parameters characterizing the effective Hamiltonian increases, so that we can expect higher-dimensional manifolds of the EP singularities to occur in the multidimensional parameter space.
Recent studies have explored higher-dimensional manifold structures of the  singularities, called {\it exceptional rings and exceptional surfaces}, and associated phenomena in topological materials and photonic systems\cite{Xu2017,Cerjan2018,Zhang2019,Okugawa19PRB,Rui2019,Cerjan2019,Budich2019,Zhou19Optica}.

While many theoretical analyses have started with the non-hermitian Hamiltonian as a fundamental time evolution generator, they have paid less attention to the dynamical origin of the dissipation processes.
Since the dissipation process is changed according to the motion of inner degrees of freedom of the subsystem as mentioned above, it becomes essential to consider the microscopic dissipation mechanism associated with each state.
Meanwhile, we have developed a theory of the {\it complex spectral analysis}\cite{Petrosky91Physica,Ordonez01PRA,Garmon2009a,Garmon2009a,Fukuta17PRA}, where the origin of the dissipation has been identified as the energy resonance in hermitian Hamiltonian for an open system.
In the complex spectral analysis, the microscopic dynamics are renormalized into the non-hermitian effective Hamiltonian written in terms of an energy-dependent self-energy function.
As a result, the eigenvalue problem becomes nonlinear in the sense that the effective Hamiltonian itself depends on its eigenvalues, whereby the dissipation dynamics are reconciled with the microscopically reversible dynamics apparent in the original Hamiltonian.

The purpose of this work is to propose a specific quantum system that exhibits the EP singularity manifold in a multidimensional parameter space and a method to observe its influence in a laboratory experiment.
For this purpose,  we consider an alkali atom doped polyacetylene as a  model system, and study the electronic Zeeman spectrum  influenced by the spin-dependent charge transfer decay.
The charge transfer process from the alkali metal to the polyacetylene conduction band has been studied mostly in terms of the induced structural changes of the interchain couplings\cite{Winokur1988,Baughman1992,Andersson1993,Sun2002}, and soliton formation\cite{Su1980PRB,Nechtschein1980,Kivelson1981,Kivelson1982,Heeger1988}.
Recent studies have indicated that the spin-orbit coupling plays an important role to determine the electronic structures even in light element materials, such as graphene and polyacetylene\cite{Konschuh2010,Yao2017a}. 
Furthermore, the crucial role of the spin-orbit coupling for the emergence of the EP manifolds has been recently clarified in Weyl metals and in photonic systems\cite{Xu2017,Cerjan2018,Cerjan2019,Zhou19Optica,Rui2019}.
We show that the EP manifolds appear in the electronic Zeeman spectrum due to the spin-dependent charge transfer decay through the spin-orbit interactions, where we see the anisotropic EP ring and isotropic EP surface.

As a second purpose, we demonstrate that coherent single electron spin resonance spectroscopy (SSESR) can be used to observe the EP singularities.
The electron spin resonance (ESR) has been widely used  to investigate the spin relaxation in a polyacetylene and molecular chain \cite{Weinberger1980,Mizoguchi1995,Kanemoto2000,Yao2017a}.
Great progress has been made in the single spin quantum control technique   in the last two decades\cite{Hanson2007,Press2008,Greve2013,Reiserer2015}, and the spatial resolution and sensitivity of the electron spin resonance spectroscopy  have now reached atomic scale precision owing to  recent advances in spin-polarized scanning tunneling microscopy or magnetic interaction with NV centers in diamond \cite{Baumann2015,Shi2015,Chen2018,Barry2020}.
It is now possible to observe a single electron spin relaxation dynamics in a single molecule with these techniques.
Here we study the coherent nonlinear single-electron spin resonance accompanied with the charge transfer decay from the alkali donor to the one-dimensional conduction band of  polyacetylene, where we apply the pump-probe pulsed excitation to obtain the nonlinear response  in addition to the static magnetic field.
It has been recognized that the nonlinear magnetic resonance spectroscopy and its multidimensional Fourier transform spectroscopy, such as two-dimensional Fourier Transform (2DFT) NMR or ESR, is a powerful tool to investigate the spin relaxation process in a molecule\cite{Tanimura1993,MukamelBook,Jonas2003}.
We show that the EP singularity as a result of  the cooperation between the Zeeman interaction and the spin-orbit coupling is well reflected in the coherent nonlinear 2DFT ESR spectrum.

It is known that the Liouville-space pathway approach in terms of the nonlinear response function is a useful tool to describe the nonlinear optical spectroscopy\cite{MukamelBook}.
In the present work, we construct a  complete basis in the Liouville space to describe the proper Liouville pathway in terms of the solutions of the complex eigenvalue problem of the total Hamiltonian, where the energy-dependence of the self-energy is essential\cite{Petrosky91Physica,Ordoez2001,Ordonez01PRA}.
We find that the time evolution following  successive pulsed excitations becomes consistent with the entropy production only when we take into account the correct analytic continuation for the Liouville states.

One of our major findings is a giant response in the 2DFT ESR associated with the EP singularity as a result of the vanishing normalization factor at the EP singularity.
The effect is known as the Petermann effect in addition to the ordinary resonant enhancement from the Purcell effect in the signal enhancement of the spontaneous emission\cite{Purcell1946,Petermann1979,Lin2016,Pick2017}.
However, what we found here is that the signal enhancement becomes orders of magnitude larger in the coherent 2DFT ESR than in the 1DFT ESR.
To our knowledge, this is the first study to reveal the giant response due to the EP singularity greatly in the nonlinear response function.

In Section \ref{Sec:Model}, we present our model of the doped polyacetylene in terms of the one-dimensional tight binding model, where the spin-orbit coupling associated with the charge transfer of the donor electron is taken into account as well as the local Zeeman effect.
In Section \ref{Sec:EVHee}, we have solved the complex eigenvalue problem of the total system Hamiltonian in terms of the Brillouin-Wigner-Feshbach projection method and obtained the complex eigenvalue spectrum in the three-dimensional  parameter space of the external magnetic field. 
We find that the EP singularities appear depending on the relative angle of the external magnetic field to the molecular axis via the spin-orbit coupling.
The coherent nonlinear 2DFT ESR spectrum as well as 1DFT ESR are formulated in terms of the Liouville space representation in Section \ref{Sec:ESR}, where the signal intensity exhibits  strong  anisotropy as the relative angle between the pump and probe directions is changed.
In addition, we find the giant response of the signal around EP singularity, indicating that this method is a useful tool to observe the single electron spin relaxation process.
We conclude in Section \ref{Sec:Conclusion} with some discussions.

%###############################################
\section{Model } \label{Sec:Model}

%%%--------------------------
\begin{figure}
\centering
\includegraphics[width=75mm,height=30mm]{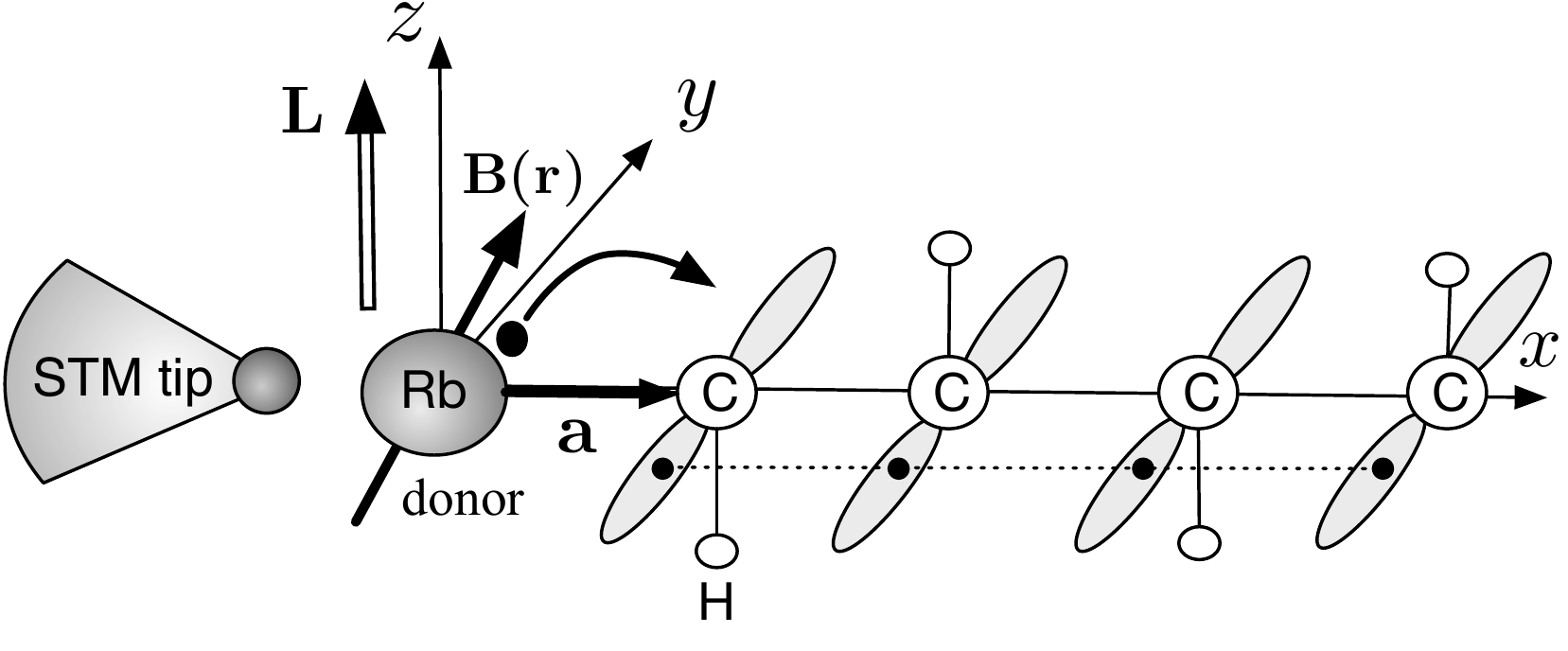}
\caption{Single-spin ESR setup for a heavy alkali donor atom binding to a one-dimensional polyacetylene. The orbital angular momentum $(\bf L)$ associated with the charge transfer is aligned in the $z$-direction, and the external magnetic field ${\bf B}({\boldsymbol r})$ acts on the spin at the donor site.}
\label{fig:Molecule} 
\end{figure}
%%%--------------------------

We consider the spin-dependent charge transfer decay of a heavy alkali atom binding to the end of a polyacetylene molecule, as shown  in Fig.\ref{fig:Molecule},  where the charge transfer occurs from a donor impurity level to the $\pi^{*}$ conduction band of the molecule.
In the present work, we take the donor site as the origin of the Cartesian coordinate axis, i.e. ${\boldsymbol r}_{{\rm D}}=0$.
In this paper, we consider the low energy Zeeman excitation, so that we neglect the effects of the $\pi$ valence band and $\sigma$-orbitals.
We describe the conduction band in terms of the one-dimensional tight-biding model with spin variables under a simple  H{\"u}ckel approximation.

Taking the energy origin at the center of the conduction band, the charge transfer Hamiltonian for a non-relativistic donor electron  is represented  by 
%--
\begin{align}
\hat H_{CT}  &= \varepsilon_D \sum_\alpha \hat c^\dagger_{0,\alpha} \hat c_{0,\alpha}+\sum_{\alpha}v\left(\hat{c}_{0,\alpha}^{\dagger}\hat{c}_{1,\alpha}+\hat{c}_{1,\alpha}^{\dagger}\hat{c}_{0,\alpha}\right)\nonumber \\
 & -J\sum_{n=1}^{N}\sum_{\alpha}\left(\hat{c}_{n+1,\alpha}^{\dagger}\hat{c}_{n,\alpha}+\hat{c}_{n,\alpha}^{\dagger}\hat{c}_{n+1,\alpha}\right) \;,\label{Hot}
\end{align}
%--
where $\hat{c}_{i,\alpha}^{\dagger}(\hat{c}_{i,\alpha})$ are the electron fermion operators with the site index $i$ and the two-spinor indices $\alpha$. 
In $\hat H_{\rm CT} $, the donor impurity atom is located at the 0-th site with the energy $\varepsilon_D$, 
the nearest neighbor transfer integral for the 1D tight binding model is $J$,  the spin-independent charge transfer from the donor to the band is represented by $v$, and $N$ is the length of the polyacetylene molecule.
With  use of the wavenumber representation 
%-- 
\begin{align}
\hat{c}_{k,\alpha}\equiv\sqrt{ 2\over N}\sum_{n=1}^{N}\sin(kn)\hat{c}_{n,\alpha}\;,\; \hat{c}^\dagger_{k,\alpha}\equiv\sqrt{ 2\over N}\sum_{n=1}^{N}\sin(kn)\hat{c}^\dagger_{n,\alpha}\;,
\end{align}
%-
and then taking the infinite limit of molecular length\cite{Fukuta17PRA}, 
the second and third terms are represented by
%--
\begin{align}
\hat{H}_{{\rm mol}}=\sum_{\alpha}\int_{0}^{\pi}dk\omega_{k}\hat{c}_{k,\alpha}^{\dagger}\hat{c}_{k,\alpha}\;,
\end{align}
and 
%--
\begin{align}
\hat{H}_{{\rm CT}}=\sum_{\alpha}\int_{0}^{\pi}dkv_{k}\left(\hat{c}_{k,\alpha}^{\dagger}\hat{c}_{0,\alpha}+\hat{c}_{0,\alpha}^{\dagger}\hat{c}_{k,\alpha}\right) \;,
\end{align}
%--
respectively, where $\omega_{k}=-2J\cos kd$ and $v_{k}\equiv\sqrt{2/\pi}v\sin kd$ with the distance between the carbon atoms $d\simeq 1.3$A used as a length unit.
In the present paper, we take $\hbar=1$.
%
%%--
%\begin{align}
%\int\left\{ g_{B} \boldsymbol{\hat{s}}({\boldsymbol r})\cdot{\bf B}({\boldsymbol r})
%+\xi_{{\rm SO}}\boldsymbol{\hat{s}} ({\boldsymbol r}) \cdot({\bf E}(\hat{{\boldsymbol r}})\times\hat{{\boldsymbol p}}) \right\} d^3{\boldsymbol r} 
%\end{align}
%%--

We consider a static local external magnetic field acting on the donor site applied by  spin-polarized STM tip, for example.
For the spin density  $\boldsymbol{s}({\boldsymbol r})$, the Zeeman interaction Hamiltonian in the  second quantized form  is given by 
%--
\begin{align}
\hat{H}_{\rm Z}({\bf B})&=-g_B\int  \boldsymbol{{s}}({\boldsymbol r})\cdot{\bf B}({\boldsymbol r})d^3{\boldsymbol r} \notag\\
&=-{g_B\over 2}\sum_{i=x,y,z}\sum_{\alpha,\beta}B_{i}(\sigma_{i})_{\alpha,\beta}\hat{c}_{0,\alpha}^{\dagger}\hat{c}_{0,\beta}\;,
\end{align}
%--
where $g_B$ is the g-factor, $B_i$ is the magnetic field on the donor site, and  $ \sigma_i\; (i=x,y,z)$ are  the Pauli matrices.
Hereafter, we rewrite $B\equiv g_B|{\bf B}|/2$, including $g_B$ into the value of $B$.

In addition, we consider the Rashba-type spin-orbit interaction\cite{SakuraiBook,Konschuh2010,Yao2017a} associated with  the charge transfer from the impurity donor to the conduction band.
The  Hamiltonian is given by
%--
\begin{align}
\hat H_{\rm SO}= \xi_{{\rm SO}} \int \boldsymbol{s} ({\boldsymbol r}) \cdot({\bf E}({\boldsymbol r})\times{\boldsymbol p}) d^3{\boldsymbol r}  \;,
\end{align}
%--
where we attribute the electric field ${\bf E}({\boldsymbol r})$ to the charge transfer polarization from the donor atom to the molecular chain, which is represented by ${\bf E}({\bf r})=e{\bf r}/\chi$ with the electric susceptibility $\chi$.
Then the spin-orbit interaction Hamiltonian reads
%--
\begin{align} \label{HSO}
\hat{H}_{{\rm SO}} & =\frac{e\xi_{{\rm SO}}}{2\chi}\int \boldsymbol{\hat{\sigma}}\cdot(\hat{{\boldsymbol r}}\times\hat{{\boldsymbol p}}) \; d^3{\boldsymbol r} \equiv g_{\rm S}\int \boldsymbol{\hat{\sigma}}\cdot{\boldsymbol {\hat l}}\; d^3{\boldsymbol r} \notag\\
&=g_{\rm S} \sum_{i=x,y,z} l_{i}(\sigma_{i})_{\alpha,\beta}\hat{c}_{0,\alpha}^{\dagger}\hat{c}_{1,\beta} +{\rm H.c.}\;,
\end{align}
%--
where $\hat{\boldsymbol l}$ is the orbital angular momentum operator.
We have evaluated the matrix element of the orbital angular momentum between the $ns$ atomic orbital of the heavy alkali donor atom and the $2p_y$ orbital of the carbon atom at the end of the molecule as
%--
\begin{align}
l_i=\<0,s|\hat l_i|1,p_y\>=\int d^{3}{\bf r}\varphi_{s}^{*}({\bf r})\hat l_{i}\varphi_{p_y}({\bf r-{\bf a}}) \;.
\end{align}
%--
We neglect the small spin-orbit couplings between the carbon $2p_y$ orbitals.
We find from the symmetry that
%--
\begin{align}\label{Lsymmetry}
l_z\neq 0\;,\; l_x=l_y=0 \;,
\end{align}
%--
as shown in Appendix \ref{AppSec:SO} \cite{Slater1954,Konschuh2010}.
In the present work, we take the $z$-axis as a quantization axis with the spin-orbit coupling parameter $L\equiv g_S l_z$ in (\ref{HSO}) as shown in Fig.\ref{fig:Molecule}.
It is seen from (\ref{HSO}) that the orbital angular momentum associated with the charge transfer plays the role of an intrinsic molecular magnetic field acting on the two-spinors, which we shall call  the {\it molecular field}, ${\bf L}=L\hat {\bf z}$.

The halfwidth of the conduction band of polyacetylene has been estimated as $2J\simeq 6$eV\cite{Heeger1988}, which we take as the energy unit.
The charge transfer strength $v$, the spin-orbit coupling strength $g_s$,  the impurity donor energy level $\varepsilon_D$ are given for a specific donor atom, while the  externally controllable parameter is the external magnetic field $\bf B$.
Usually these parameter values satisfies $2J \gg v\gtrsim g_S$.
In this work, we take $\varepsilon_D=0$, and we make a brief comment on  the effect of $\varepsilon_D$ in   Appendix \ref{AppSec:GeoHtot}.
Since we focus on a rapid motion of the donor electron, we neglect the effect of the slow motion of the structural change of the polyacetylene molecular backbone.
In the next section, we study the complex spectra of the {\it total} Hamiltonian in terms of the external magnetic field $\bf B$-parameter space, and reveal the EP singularities in  this space.

%#################################################
\section{Complex eigenvalue problem of the Hamiltonian and the EP singularities}
\label{Sec:EVHee}

In this section we solve the complex eigenvalue problem of the total Hamiltonian.
We first derive the effective Hamiltonian in terms of the Brillouin-Wigner-Feshbach (BWF) projection method, and obtain the complex eigenspectrum of the Hamiltonian. 

The complex eigenvalue problem of the total Hamiltonian for the parameter $\bf B$ reads
\cite{Petrosky91Physica,Tanaka06PRB,Fukuta17PRA}:
%---
\begin{subequations}\label{HtotEVP} 
\begin{align}
 & \hat{H}_{{\rm tot}}({\bf B})|\phi_{\xi}({\bf B})\>=z_{\xi}({\bf B})|\phi_{\xi}({\bf B})\>\;,\\
 & \<\tilde{\phi}_{\xi}({\bf B})|\hat{H}_{{\rm tot}}({\bf B})=z_{\xi}({\bf B})\<\tilde{\phi}_{\xi}({\bf B})|\;,
\end{align}
\end{subequations} 
%--
where the right- and the left-eigenstates, $|\phi_{\xi}({\bf B})\>$ and $\<\tilde{\phi}_{\xi}({\bf B}|$,
respectively, share the same eigenvalue $z_{\xi}({\bf B}) $.
For the current problem, we restrict ourselves to a single-particle vector space consisting of $\{ |n,\alpha\>  \; (n=0, 1, \cdots ; \alpha=\pm )\}$, where  $|n,\alpha\>\equiv c^\dagger_{n,\alpha}|{\rm vac}\>$ and $|{\rm vac}\>$ denotes the electron vacuum.

We shall reduce the eigenvalue problem of the total Hamiltonian into the $\{|0,+\>, |0,-\>\}$-space in terms of the projection operators
%--
\begin{align}
\hat P_{0}=|0,+\>\<0,+| +  |0,-\>\<0,-| \;,\; \hat Q_{0}=1-\hat P_{0} \;.
\end{align}
%--
 Taking into account (\ref{Lsymmetry}), we have derived the effective Hamiltonian matrix  in the $\hat P_0$ subspace as
%--
\begin{align}\label{Heff} 
H_{\rm eff}({\bf B};z)  
 & =-\begin{pmatrix} B_{z} & B_{x}-iB_{y}\\
B_{x}+iB_{y} & -B_{z}
\end{pmatrix} \notag\\
&
+\Sigma(z)\begin{pmatrix}(v+L)^{2} & 0\\
0 & (v-L)^{2}
\end{pmatrix}\;.
\end{align}
%--
The derivation of the effective Hamiltonian is shown in Appendix \ref{AppSec:Heff}.
The microscopic dissipation effect to the continuum is renormalized into $H_{\rm eff}$ by the self-energy function $\varSigma(z)$ given by a Cauchy integral calculated as
%--
\begin{align}\label{Self}
\varSigma^{\pm}(z) & =\frac{2}{\pi}\int_{0}^{\pi}dk\frac{\sin^{2}k}{(z-\omega_{k})^{\pm}}=\frac{1}{2}\left(z\mp\sqrt{z^{2}-4}\right)\;.
\end{align}
%--
The self-energy has a branch cut in the region  $[-2:2]$ on the real z-axis, where the sign in the integral indicates the analytic continuation from the upper($+$) and lower($-$)-half complex $z$-plane corresponding to the retarded and advanced Green's function, respectively\cite{Petrosky91Physica}.
By the definition of (\ref{Self}), the self-energy is an odd function in terms of $z$ as
%--
\begin{align}\label{SelfOdd}
\varSigma^\pm(-z)=-\varSigma^\mp(z)
\end{align}
%--
reflecting the microscopic interaction with the continuum\cite{Tanaka2020}. 
As will be shown later, this relation is crucial to ensure the symmetry ({\it chiral symmetry}) to give rise to the EP ring structure. 

We  emphasize that the complex eigenvalues of the effective Hamiltonian are the same as those of the total hermitian Hamiltonian only when we take into account the energy-dependence of the self-energy\cite{Petrosky91Physica,Ordoez2001,Tanaka06PRB}.
As a result, the complex eigenvalues appear as complex-conjugate pairs, if they exist, corresponding to the resonance and anti-resonance eigenstates, which we shall call {\it the dynamical symmetry}.
The solution pairs thus obtained are the fundamental building blocks to construct the Liouville space basis used in the formulation of the coherent ESR spectroscopy as shown in the next section.

It is seen that the Zeeman effect on the two-spinor at the donor site, the first term in (\ref{Heff}), is modified by the dissipation effect of the second term.
We  emphasize that the spin-orbit coupling $L$ effectively enhances or suppresses the charge transfer effect $v$ when the spin direction is parallel or antiparallel to the molecular field $\bf L$, respectively: The charge transfer decay process becomes spin-dependent due to the spin-orbit coupling.

The complex eigenvalues of the effective Hamiltonian are obtained by solving  the characteristic equation 
%--
\begin{align}\label{CharaEq}
&\eta^\pm(z;{\bf B})\equiv  \left\{ z-B_{z}-(v-L)^{2}\varSigma^{\pm}(z)\right\} \notag\\
&\times\left\{ z+B_{z}-(v+L)^{2}\varSigma^{\pm}(z)\right\}   -(B_{x}^{2}+B_{y}^{2})=0\;.
\end{align}
%--
Squaring out the root in the self-energy, we find  the characteristic equation is a fourth-order polynomial equation, yielding
 two complex conjugate pair solutions in the resonant regime, as mentioned above, and four real solutions in the off-resonant regime.  
%We have numerically solved the characteristic equation (\ref{CharaEq}) and  found that  there are two solutions $z_\xi^\pm({\bf B})$ for each analytic continuation corresponding to the two-spinors.
With  use of the spherical coordinate  ${\bf B}=(B\sin\theta_B\cos\phi_B,B\sin\theta_B\cos\phi_B,B\cos\theta_B)$, one can immediately see from (\ref{CharaEq}) that the solutions depend only on the polar angle $\theta_B$ as well as the field strength $B$.

%--------------------------------------
\begin{figure}[ht]
\begin{center}
\includegraphics[width=80mm,height=40mm]{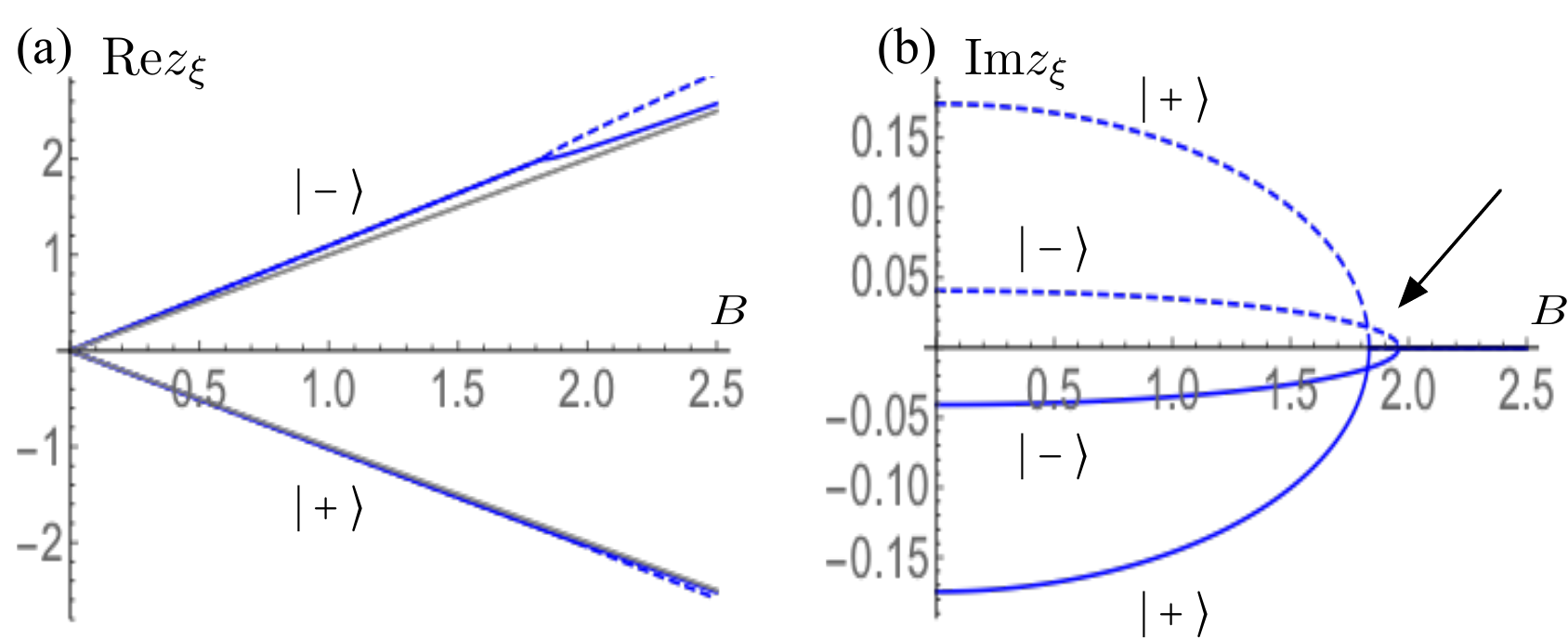}
\caption{Eigenvalues of the effective Hamiltonian as a function of ${\bf B}$ parallel to ${\bf L}$ taken by $\theta_{B}=0$ for  $v=0.3$ and $L=0.1$.
(a) Real part and (b) imaginary part. 
The solid and dashed lines in (b) correspond to the resonance and anti-resonance state solutions in the resonant regime, respectively.
In the off-resonant regime, the solid and dashed lines correspond to the  bound and virtual bound states  shown in (a).
The gray lines in (a) are guidelines for the unperturbed energies of the two-spinors. }
\label{fig:E0} 
\end{center}
\end{figure}
%----------------

First we show the results of the complex eigenvalues when the external magnetic field is parallel to the molecular field, i.e. ${\bf B}\parallel  {\bf L}=L \hat{\boldsymbol z}$, i.e. $B_x=B_y=0$.
It is immediately seen from (\ref{Heff}) that the two-spinor states are decoupled in this case, and  the characteristic equation (\ref{CharaEq}) becomes a product of the two independent equations associated with $|+\>$ and $|-\>$ spin states.
We show in Fig.\ref{fig:E0} the complex eigenvalues of the effective Hamiltonian as a function of the magnetic field strength $B$ for $\theta_B=0$ $({\bf B}\parallel {\bf L})$, where the real and imaginary parts of the eigenvalues are shown in Fig. \ref{fig:E0}(a) and (b), respectively.
The other parameters are taken as $v=0.3$ and $L=0.1$. 
The solutions for the different analytic continuations of the self-energies of $\varSigma^+(z)$ and $\varSigma^-(z)$  are depicted by the solid and dashed lines, respectively.
The solutions for $\varSigma^+(z)$ and $\varSigma^-(z)$ correspond to the resonance and anti-resonance states in the resonant regime, respectively.

As seen in Fig.\ref{fig:E0}(a), the real parts of the eigenvalues are split by the Zeeman effect: The real part of the energy corresponding to the parallel (antiparallel) spin states to $\bf B$ linearly decreases (increases) with $B$.
Although this behavior is similar to the ordinary Zeeman effect on an isolated two-spinor, it should be noted that the eigenstates in the present case are resonance (anti-resonance) states which do not belong to the Hilbert space\cite{Petrosky91Physica}.
Indeed, the eigenstates have an imaginary part of the eigenvalues as long as the Zeeman split energies are in resonance with the conduction band, as shown in Fig.\ref{fig:E0}(b).
As mentioned above, the charge transfer decay couplings become spin-dependent due to the spin-orbit coupling, so that in this case the parallel spin state $|+\>$ has a larger decay rate than the anti-parallel spin state $|-\>$.
 
 As $B$ increases and the Zeeman split energies come close to the conduction band edge $B\simeq 2$ , the resonance and anti-resonance states coalesce at the EP singularity depicted by the arrow in Fig.\ref{fig:E0}(b).
 As $B$ further increases, they are split into the bound and virtual-bound states\cite{Garmon2015a,Garmon2017a}. 
The EP singularities  near the band edge $B\simeq 2$ appear as a result of the energy resonance singularity between the discrete Zeeman states and the conduction continuum irrespective to the direction of $\bf B$.

%--------------------------------------
\begin{figure}[t]
\begin{center}
\includegraphics[width=80mm,height=40mm]{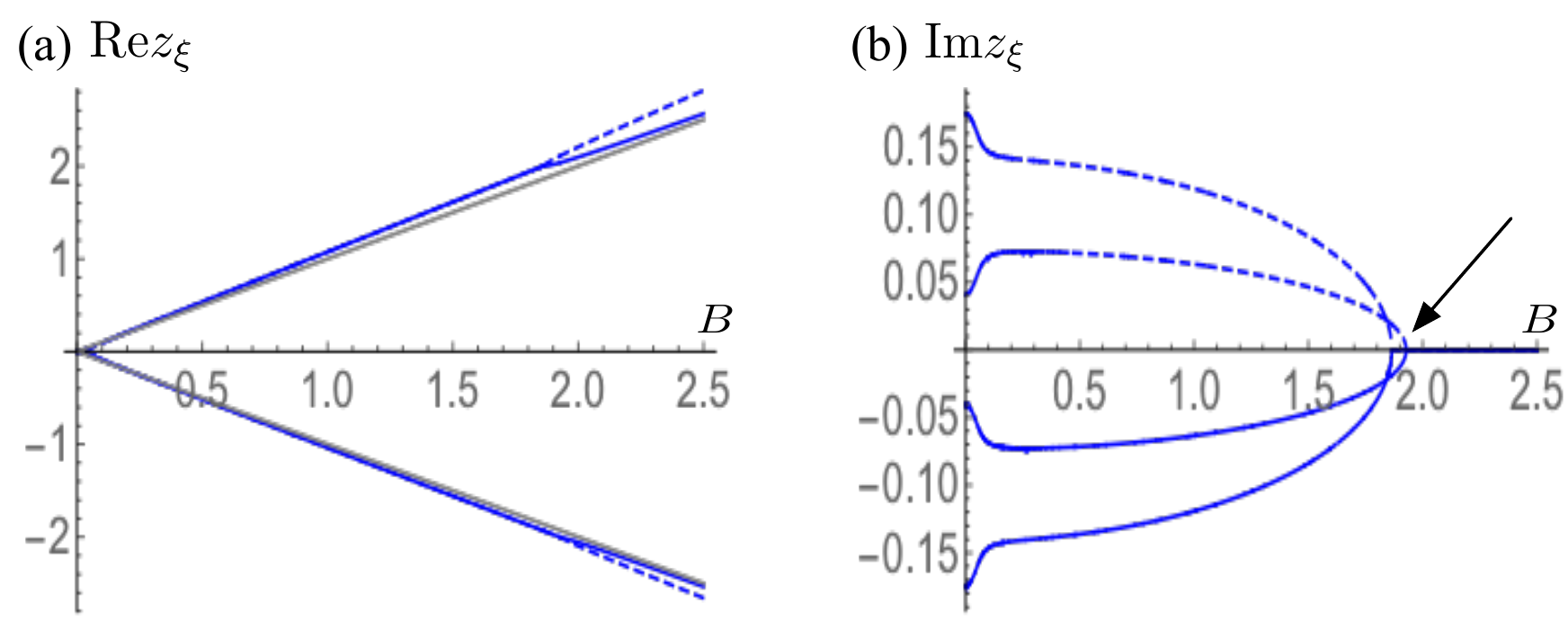}
\caption{Eigenvalues of the effective Hamiltonian as a function of ${\bf B}$ parallel to ${\bf L}$ taken by $\theta_{B}=\pi/3$ for  $v=0.3$ and $L=0.1$.
(a) Real part and (b) imaginary part. 
The solid and dashed lines have the same meaning as in Fig.\ref{fig:E0}. }
\label{fig:EPmiddle} 
\end{center}
\end{figure}
%----------------

\begin{figure}[t]
\begin{center}
\includegraphics[width=85mm,height=50mm]{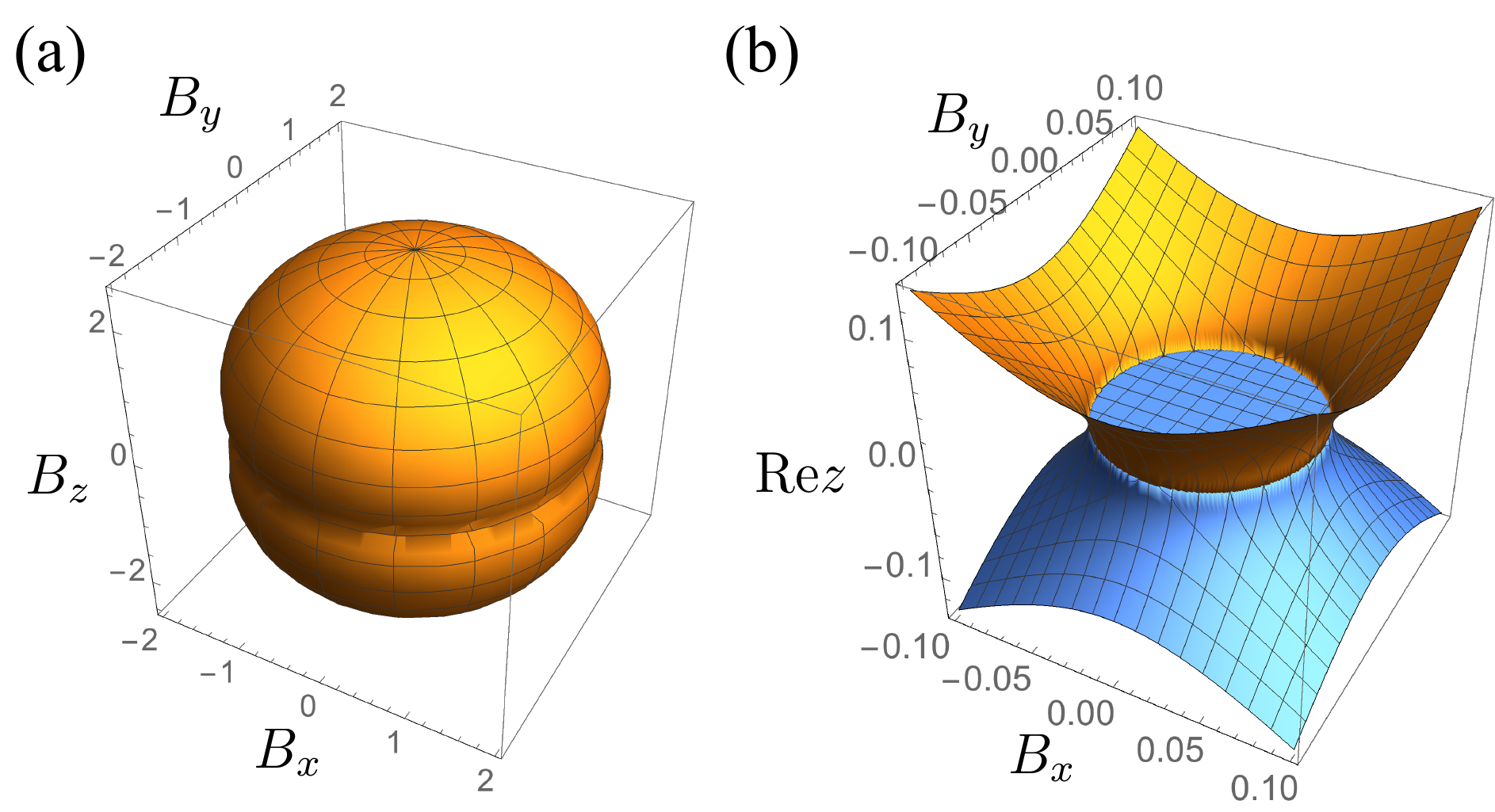}
\caption{The EP manifolds for $v=0.3$ and $L=0.1$. (a) EP surface in the $\bf B$-space. (b) Real part of the complex eigenvalues in the $B_x$-$B_y$ plane, where EP ring appears at $|B|=B_{\rm EP,-}$.}
\label{fig:EPmanifold} 
\end{center}
\end{figure}

When we tilt $\theta_B$ away from the polar axis of ${\bf L}$, the two-spin states become coupled.
We show the results of $\theta_{B}=\pi/3$ in Fig.\ref{fig:EPmiddle}. 
The effect of $\theta_{B}$ is prominent in the imaginary part of the eigenvalues; As $\theta_{B}$ increases
from $0$, the gap of the imaginary parts of the complex eigenvalues of the two resonance states or the two anti-resonance states
is narrowed, while the values at ${\bf B}=0$ are unchanged.
We see the EP singularities at the band edge still remain as indicated by the arrow.
Since this EP singularity appears as a result of the energy resonance between the discrete Zeeman  states and the conduction continuum irrespective of the polar angle $\theta_B$, the EP singularity forms an EP surface in the $\bf B$-space as shown in  Fig.\ref{fig:EPmanifold}(a).

%--------------------------------------
%----------------

As mentioned above and shown in Appendix \ref{AppSec:GeoHtot}, the total system Hamiltonian possesses the {\it chiral symmetry}, where \eqref{AppEq:HtotSym}  holds when $B_z=0$. 
Since the effective Hamiltonian is derived from the total Hamiltonian and the microscopic spin-dependent charge transfer decay is taken into account by the energy-dependent self-energy,   the effective Hamiltonian also possesses the chiral symmetry: We found that at $B_z=0$, owing to  (\ref{SelfOdd}), the effective Hamiltonian at $B_z=0$ satisfies
%--
\begin{align}\label{HeffSymp}
\sigma_z\hat H_{\rm eff}(z)  \sigma_z=-\hat H_{\rm eff}(-z)  \;,
\end{align}
%--
with the Pauli matrix $\sigma_z$.
Note the correspondence with \eqref{AppEq:HtotSym}, which indicates the derived effective Hamiltonian maintains the chiral symmetry of the total Hamiltonian.

As is the same with \eqref{AppEq:HtotSymEV}, it follows from  (\ref{HeffSymp}) that
%--
\begin{align}
\hat H_{\rm eff}(z)|\psi\>=z|\psi\>  \implies  \hat H_{\rm eff}(-z)\ \sigma_z|\psi\>=-z  \sigma_z|\psi\> \;.
\end{align}
%--
This relation ensures the appearance of pairs of solutions with opposite signs, in addition to the complex conjugate pair  solutions due to the dynamical symmetry as mentioned just below (\ref{SelfOdd}).
As a result of the combination of the dynamical symmetry and the chiral symmetry, the complex eigenvalues are obtained as a set of  four solutions in the form   
%--
\begin{align}\label{foursolutions}
\text{resonance}
\begin{cases}
 z_1=\alpha-i\beta \;, \\
\bar z_1=-\alpha-i\beta \;,
\end{cases}
\text{anti-resonance}
\begin{cases}
 z_1^*=\alpha+i\beta \;,\\
 \bar z_1^*=-\alpha+i\beta \;,
\end{cases}
\end{align}
%--
with $\alpha,\beta>0$.
Here we denote the upper bar and asterisk as taking the opposite sign of the real and imaginary parts, respectively, i.e., $-z=\bar z^*$.
The advantage of our theory is that we can deal with the two different types of  symmetry breaking, the dynamical symmetry and the chiral symmetry breaking, in a unified manner.
The effective Hamiltonian maintains the symmetry of the original Hamiltonian by correctly taking into account the energy-dependence of the self-energy.

The characteristic equation is then given by
%--
\begin{align}\label{chracterEq}
& \{(v-L)^{2}-1\}\{(v+L)^{2}-1\}z^{4} \notag\\
&+[4(v^{2}+L^{2})^{2}-2(v^{2}-L^{2})^{2}(v^{2}+L^{2}+1)\nonumber \\
 & -B^{2}\{(v^{2}-L^{2})^{2}-2(v^{2}+L^{2}-1)\}]z^{2} \notag\\
 &+\{B^{2}+(v^{2}-L^{2})^{2}\}^{2}=0\;,
\end{align}
%--
where the quartic polynomial equation in $z$  reduces to a quadratic polynomial equation  of $z^2$.
This form is the same as  studied in the different context of parametric amplification in the dynamical Casimir system\cite{Tanaka2020,Tanaka2020PTEP}, where it was shown that the symplectic symmetry of the system is crucial to cause the EP singularity.

The EP singularities are determined by solving the discriminant equation of (\ref{chracterEq}): 
%--
\begin{align}
D(B;v,L)=B^{4}+4(v^{2}+L^{2}-1) B^2+16v^{2}L^{2}=0\;,
\end{align}
%--
which gives the EP singularities at $B=B_{{\rm EP},\pm}$, where
%--
\begin{align}
&(B_{{\rm EP},\pm})^{2}\notag\\
&=2\left\{(1-v^{2}-L^{2})\pm\sqrt{(1-v^{2}-L^{2})^2-4v^2 L^2}\right\} >0\;.
\end{align}
%--
The EP singularities of $B_{\rm EP,+}$  and $B_{\rm EP,-}$ correspond respectively to the dynamical symmetry breaking and the chiral symmetry breaking points. 
Taking into account that  $2J=1\gg v \gtrsim L$, we approximately evaluate $B_{{\rm EP},-}\simeq 2 v L$ and  $B_{{\rm EP},+}\simeq 2$.
Since the chiral symmetry breaking point $B_{\rm EP,-}$ appears only in the $B_z=0$ plane, it forms an EP ring structure, while  $B_{\rm EP,+}$ corresponds to the EP surface.

%--------------------------------------
\begin{figure}[t]
\begin{center}
\includegraphics[width=80mm,height=80mm]{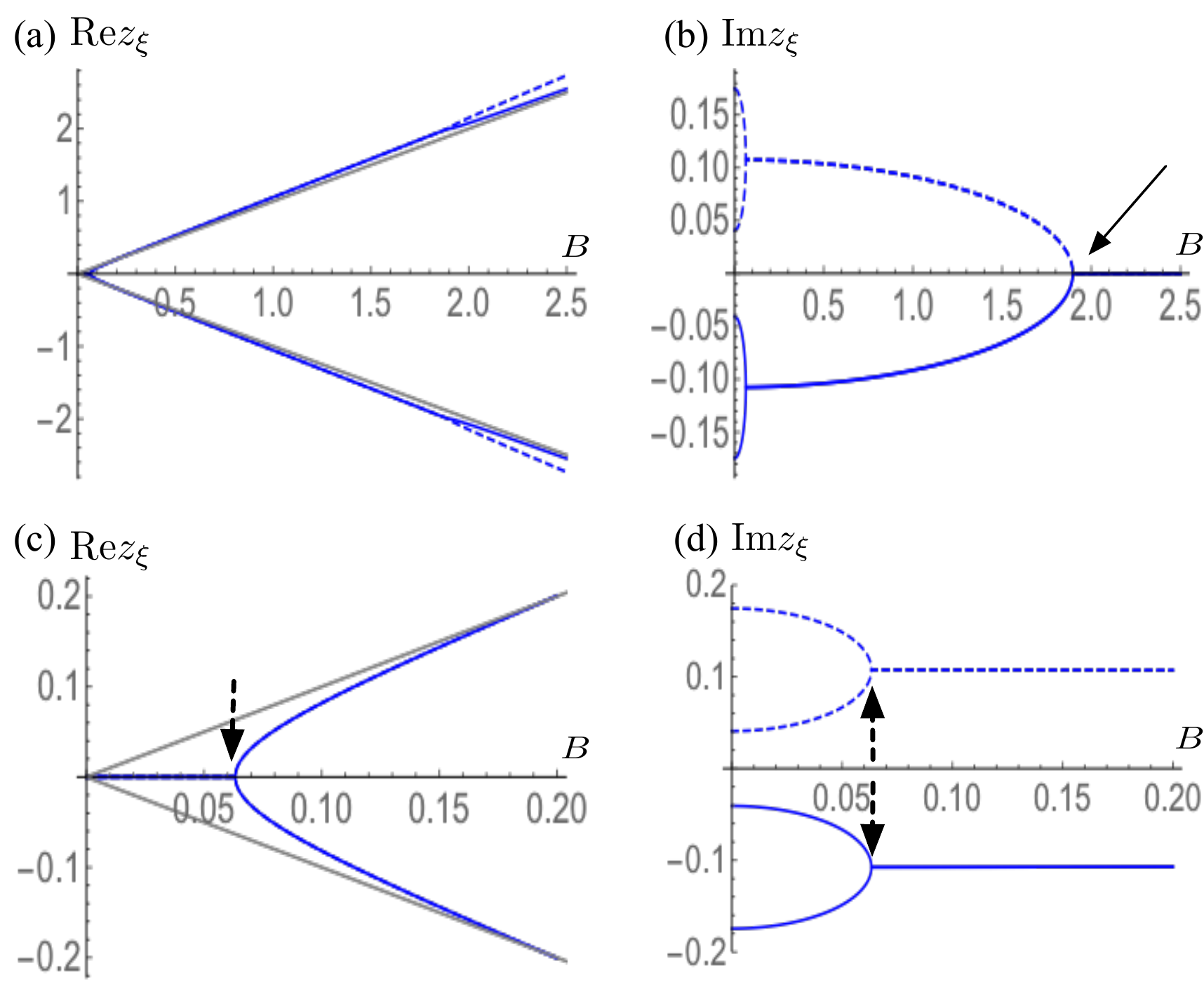}
\caption{Eigenvalues of the effective Hamiltonian as a function of ${\bf B}$ perpendicular to ${\bf L}$ taken by $\theta_{B}=\pi/2$ $(B_z=0)$  for  $v=0.3$ and $L=0.1$.
(a) Real part and (b) imaginary part. 
The gray lines in (a) are guidelines for the unperturbed energies.
 The spectra of the real and imaginary parts are shown in the expanded horizontal scale in (c) and (d), respectively.
 The arrows indicate $B_{\rm EP,+}$ in (b) and $B_{\rm EP,-}$ in (c) and (d).
 }
\label{fig:EPring} 
\end{center}
\end{figure}
%----------------

We show in Fig.\ref{fig:EPring} the complex eigenvalues of the effective Hamiltonian at $B_z=0$, where the real and imaginary parts are depicted in (a) and (b), respectively.
The arrow in (b) indicates the dynamical symmetry breaking point at $B_{\rm EP,+}$. 
We show the real and imaginary parts in Fig.\ref{fig:EPring}(c) and (d), respectively, in the expanded scale around $B_{\rm EP,-}$, where the arrows indicate the chiral symmetry breaking points.
Since this EP singularity appears only at $B_z=0$, we see the EP ring structure in the $B_x$-$B_y$ plane as shown in  Fig.\ref{fig:EPmanifold}(b).

Here we compare our results with a phenomenological parity-time (PT) Hamiltonian model which has been used to study the exceptional rings and surfaces\cite{Xu2017,Cerjan2018,Cerjan2019,Zhou19Optica,Rui2019}.
In the phenomenological PT model, the dissipation term is represented by an imaginary  constant, such as $i\gamma$.
Indeed, when we approximate $\Sigma(z)$ by $-i$ in (\ref{Self}), then the effective Hamiltonian is approximated by 
%--
\begin{align}
\hat H_{\rm eff}({\bf B})  
 \simeq-i(v^2+L^2)-\begin{pmatrix} B_{z}-2ivL  & B_{x}-iB_{y}\\
B_{x}+iB_{y} & -B_{z}+2ivL
\end{pmatrix} \;.
\end{align}
%--
For $B_z=0$, 
%--
\begin{align}
\hat  H_{\rm eff}({\bf B})=-i(v^2+L^2)+\hat H_{\rm PT}({\bf B}) \;,
\end{align}
 where $\hat H_{\rm PT}({\bf B})$ is a phenomenological PT Hamiltonian 
%--
\begin{align}
\hat H_{\rm PT}({\bf B})=
\begin{pmatrix} -2ivL  & B_{x}-iB_{y}\\
B_{x}+iB_{y} & 2ivL
\end{pmatrix} \;.
\end{align}
%--
The eigenvalues are given by
%--
\begin{align}
z_{\rm PT,\pm}=-i(v^2+L^2)\pm\sqrt{(B_x^2+B_y^2)-4 v^2 L^2} \;.
\end{align}
%--
Therefore, this phenomenological treatment also explains the emergence of the EP ring at $B_{\rm PT,EP,-}=2 vL$ for the $B_z=0$ case.
However, this phenomenological model gives only the resonance eigenstates, but not the antiresonance eigenstates which is necessary to construct the Liouville space basis in the  calculation of the SSESR.
On the other hand, in the present work, we have obtained a set of four solutions (\ref{foursolutions}) by taking into account the energy-dependent self-energy.

%###########################################
\section{Nonlinear coherent SSESR spectrum in the Liouville space}\label{Sec:ESR}

The EP singularities revealed in the preceding section  significantly influence  the ultrafast single-spin resonance spectroscopy.
We consider the pump-probe single-spin free induction decay for the present system.
In the pump-probe process, an electron with spin parallel to the external static magnetic field ${\bf B}$ is transferred from the STM tip to the donor site by the pump electric field.
The interaction with the pump field is described by
%--
\begin{align}
\hat V(t)={\cal E}^*(t) |S,B_+\>\<0,B_+|+{\cal E}(t) |0,B_+\>\<S,B_+| \;,
\end{align}
%--
where $|S,B_+\>$ denotes an electronic state with the spin parallel to the external magnetic field at the STM tip, and ${\cal E}(t)$ (${\cal E}^*(t) )$ denotes the electric pump-pulse field. 

After some delay time $t_1$, we introduce the probe microwave magnetic field to induce the transition between the Zeeman split states.
The interaction  of the electronic spin and the probe magnetic field ${\bf B}_1({\boldsymbol r},t) $ which is localized on the donor site is given by
%--
\begin{align}
\hat W(t)=-g_B\int \hat{\bs  s}({\boldsymbol r})\cdot {\bf B}_1({\boldsymbol r},t) d^3 {\boldsymbol r} \;.
\end{align}
%--
The probe magnetic impulsive pulse for the free induction decay  is represented by
%--
\begin{align}\label{Probe}
{\bf B}_1({\bs r}_D,t) ={{\cal B}_1({\bs r}_D},t-t_1)+ {\cal B}^*_1({\bs r}_D,t-t_1) \;,
\end{align}
%--
where ${\cal B}_1({\bs r},t) $ and ${\cal B}^*_1({\bs r},t) $ are the fields which are assumed to be localized at the donor site.

%%--------------------------------------
%\begin{figure}[t]
%\begin{center}
%\includegraphics[width=60mm,height=20mm]{Pulse}
%\caption{Phase-locked pulse trains. }
%\label{fig:Pulse} 
%\end{center}
%\end{figure}
%%----------------
%

\begin{figure}[htbp]
\begin{center}
\includegraphics[width=80mm,height=50mm]{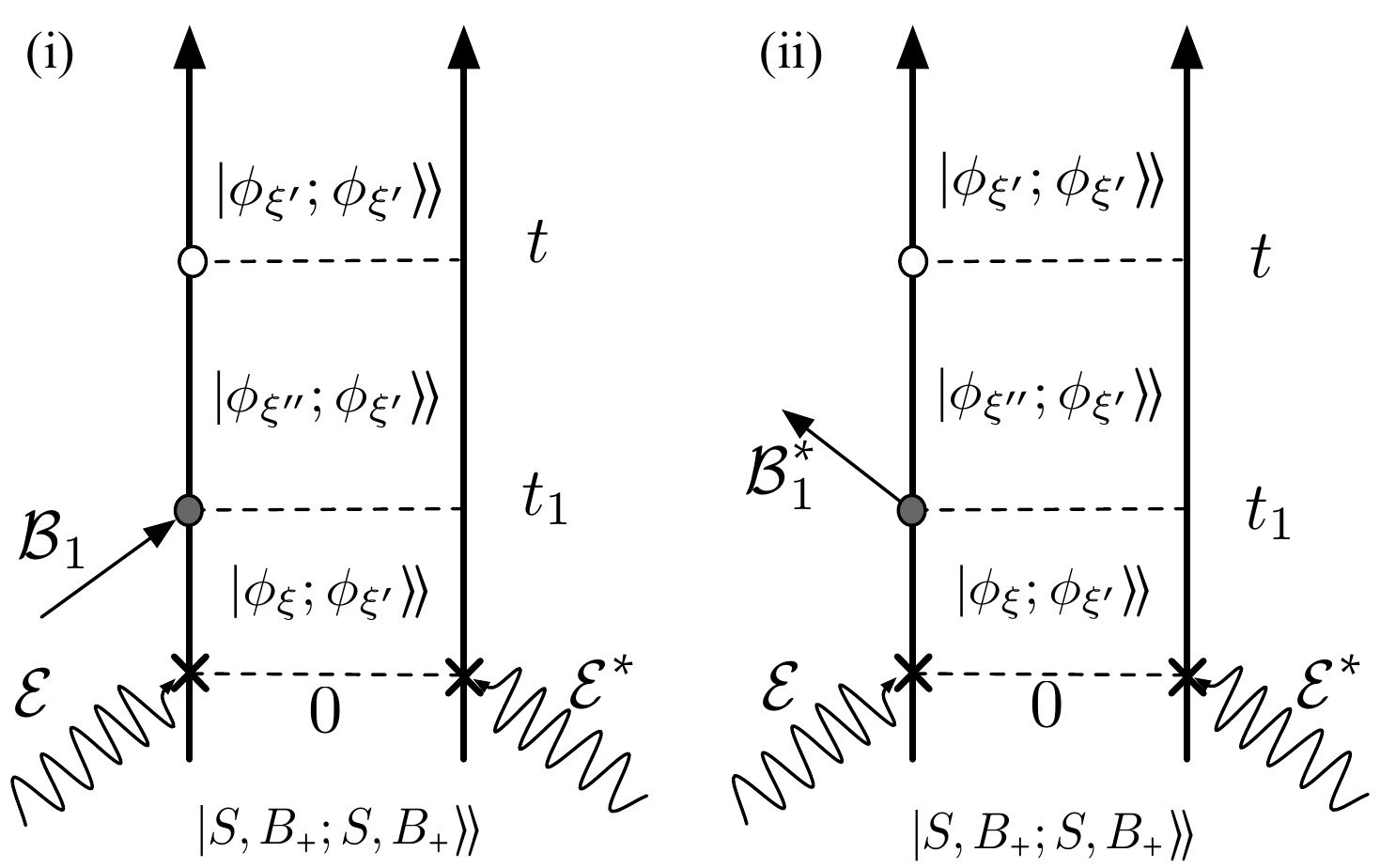}
\caption{Double sided Feynman diagrams for the coherent ESR spectroscopy.
The wavy line indicates the interaction with the pump excitation pulse, and the arrow indicates the interaction with the probe magnetic pulse field. }
\label{fig:Diagram} 
\end{center}
\end{figure}

We measure the induced spin polarization at the donor site  in terms of the Liouville space representation\cite{MukamelBook,Tanaka01PRA01,Tanaka2002,Tanaka2002a,Tanaka03JCP,Tanaka2003a,Tanaka2003}.
The third order induced spin polarization is given by
%--
\begin{align}\label{srtLiouville}
{\bs s}({\bs r}_D,t)&=(-i)^3 \int_{t_0}^t d\tau_3 \int_{t_0}^{\tau_3} d\tau_2 \int_{t_0}^{\tau_2} d\tau_1 \<\!\< \hat{\bs s}({\bs r}_D)|e^{-i {\cal L}_{\rm tot}(t-\tau_3)} \notag\\
&\times  {\cal W}(\tau_3) e^{-i {\cal L}_{\rm tot} (\tau_3-\tau_2)}
 {\cal V}(\tau_2) e^{-i {\cal L}_{\rm tot} (\tau_2-\tau_1)} \notag\\
&\times  {\cal V}(\tau_1) e^{-i {\cal L}_{\rm tot} (\tau_1-t_0)} |\rho (0)\>\!\> \;,
\end{align}
%--
where ${\cal L}_{\rm tot}$ is the Liouvillian superoperator in the Liouville space defined by ${\cal L}_{\rm tot}\cdot \equiv [\hat H_{\rm tot},\cdot]$ , and $|\rho(0)\>\!\>$ is the initial state density operator at an initial time $t_0$:
%--
\begin{align}
|\rho(0)\>\!\> =|S,B_+;S,B_+\>\!\> \;.
\end{align}
%--
In (\ref{srtLiouville}), the curly operators denote the  superoperators corresponding to the Hilbert space operators.
It is essential to describe the coherent spectroscopy in terms of the Liouville space representation in order to register the correct decay processes\cite{Ordonez01PRA,Ordoez2001}.
We have shown in Fig.\ref{fig:Diagram} the double-sided Feynman diagram corresponding to the Liouville-space pathway of (\ref{sirt}).

%--------------------------------------
%----------------

In the present work, we consider the impulsive pulses of the electric pump and magnetic probe fields as
%--
\begin{align}
{\cal E}(t), {\cal E}^*(t), {\cal B}_1({\bs r}_D,t), {\cal B}_1^*({\bs r}_D,t) \propto\delta(t) \;.
\end{align}
%--
We represent (\ref{srtLiouville}) in terms of the Liouville space basis defined by (\ref{AppEq:Lbasis}), where we have utilized the complex eigenstates of the total Hamiltonian defined by (\ref{HtotEVP}).
The explicit representation of the complex eigenstates of the total Hamiltonian are given in Appendix \ref{AppSec:Heff}.
Applying \eqref{AppEq:LiouvillComp} to \eqref{srtLiouville}, we have obtained the induced spin polarization component as  
%--
\begin{align}\label{sirt}
s_i({\bs r}_D,t)&\simeq(-i)^3\sum_{\xi,\xi',\xi''}e^{-i \Delta_{\xi'',\xi'}(t-t_1)} e^{-i \Delta_{\xi,\xi'} t_1} \notag\\
&\times\<\!\< \hat{ s}_i({\bs r}_D)|\phi_{\xi''};\phi_{\xi'}\>\!\>\<\!\<\tilde\phi_{\xi''};\tilde\phi_{\xi'}| {\cal W}(t_1)  |\phi_{\xi};\phi_{\xi'}\>\!\>  \notag\\
&\hspace {3cm} \times\<\!\<\tilde\phi_{\xi};\tilde\phi_{\xi'}|0,B_+;0,B_+\>\!\>  \;,
\end{align}
%--
 where $\xi,\xi',\xi''$ denote the discrete resonance eigenstates and we neglect the contributions from the continous states, assuming weak couplings of $v$ and $L$.

With  use of (\ref{AppEq:Lbasis}), we obtain the nonlinear response function for $s_i({\bs r}_D,t)$ to the probe field component ${\bf B}_{1,j}$  as
%-
\begin{align}
&F_{i,j}(t_2,t_1)= 4 g_B  \sum_{\xi,\xi',\xi''} {\rm Im}\left[ e^{-i\Delta_{\xi'',\xi'} t_2} e^{-i\Delta_{\xi,\xi'}t_1} \right.
 \notag\\
 &\left. \times\< 0,B_+|\tilde\phi_{\xi'} \> \<\phi_{\xi'} | \hat  s_i({\bs r}_D)|\phi_{\xi''}\>  \<\tilde\phi_{\xi''} | \hat  s_j({\bs r}_D)  |\phi_{\xi}\> \<\tilde\phi_{\xi} |0,B_+\> \right] \;,
 \end{align}
%--
where  $t_2\equiv t-t_1$.
By the two-dimensional Fourier transformation, we obtain the 2DFT SSESR spectrum in the frequency domain as
%--
\begin{subequations}\label{chi}
\begin{align}
&{\cal \chi}_{i,j}(\omega_2,\omega_1)\equiv  4 g_B    \notag\\
&\times \sum_{\xi,\xi',\xi''} {\rm Im}\left[ \int_0^\infty dt_2 \int_0^\infty dt_1 e^{i\omega_2 t_2+i\omega_1 t_1} e^{-i\Delta_{\xi'',\xi'} t_2-i\Delta_{\xi,\xi'}t_1} \right. \notag\\
&\left. \times\< 0,B_+|\tilde\phi_{\xi'} \> \<\phi_{\xi'} | \hat  s_i({\bs r}_D)|\phi_{\xi''}\>  \<\tilde\phi_{\xi''} | \hat  s_j({\bs r}_D)  |\phi_{\xi}\> \<\tilde\phi_{\xi} |0,B_+\> \right]  \notag\\
&= -4 g_B \sum_{\xi,\xi',\xi''} {\rm Im}\left[ {1 \over \omega_2-\Delta_{\xi'',\xi'}} {1 \over \omega_1- \Delta_{\xi,\xi'}}  \right.
 \notag\\
 &\left. \times\< 0,B_+|\tilde\phi_{\xi'} \> \<\phi_{\xi'} | \hat  s_i({\bs r}_D)|\phi_{\xi''}\>  \<\tilde\phi_{\xi''} | \hat  s_j({\bs r}_D)  |\phi_{\xi}\> \<\tilde\phi_{\xi} |0,B_+\> \right]  \\
& \equiv  -4g_B \sum_{\xi,\xi',\xi''} {\rm Im}\left[\chi_{i,j}^{(\xi,\xi',\xi'')}(\omega_2,\omega_1) \right]
 \end{align}
 \end{subequations}
%--
where the spectrum of $\omega_1$ indicates the spin-polarization coherence with the pump process, while that of $\omega_2$ indicates the induced polarization by the probe pulse.
The selection rule for the allowed transition under the far-off resonant case is given by
%--
\begin{align}\label{Selection}
\< 0,B_+| \hat  s_i({\bs r}_D)  \hat  s_j({\bs r}_D)  |0,B_+\> 
=\begin{cases}
 {i\over 2} \< 0,B_+| \hat  s_k({\bs r}_D)  |0,B_+\> \epsilon_{ijk}  \;,\\
 \delta_{i,j}
 \end{cases} 
\end{align}
%--
where $\epsilon _{ijk}$ is the Levy-Civita symbol.
It is seen from  (\ref{chi}a) that the quantum coherence between the resonance states  is  reflected in the cross correlation of  $\omega_1$ and $\omega_2$ in the 2DFT SSESR spectrum\cite{Tanimura1993,Jonas2003}.

\begin{widetext}
We have derived the explicit form of the discrete eigenstates of the total Hamiltonian in terms of the projection method given by (\ref{AppEq:phixi}) and (\ref{AppEq:leftphi}) as
%--
\begin{subequations}
\begin{align}
|\phi_\xi\>&={1\over {\cal N}_\xi^{1/2} }\left[  \left\{ |0,+\> + (v+L) \sqrt{2\over \pi}\int_0^\pi dk {\sin k \over (z-\omega_k)_{z_\xi}^+}|k,+\> \right\} \right. \notag\\
&\left. + {B_x+ i B_y \over  B_z+(v-L)^2 \Sigma^+(z_\xi)-z_\xi }  \left\{ |0,-\> + (v-L) \sqrt{2\over \pi}\int_0^\pi dk {\sin k \over (z-\omega_k)_{z_\xi}^+}|k,-\> \right\}  \right]    \;,\\
\<\tilde\phi_\xi| &={1\over {\cal N}_\xi^{1/2} } \left[  \left\{ \<0,+ | + (v+L) \sqrt{2\over \pi}\int_0^\pi dk {\sin k \over (z-\omega_k)_{z_\xi}^+}\<k,+| \right\} \right. \notag\\
&\left. +{B_x- i B_y \over  B_z+(v-L)^2 \Sigma^+(z_\xi)-z_\xi} \left\{ \<0,-| + (v-L) \sqrt{2\over \pi}\int_0^\pi dk {\sin k \over (z-\omega_k)_{z_\xi}^+}\<k,-| \right\} \right] \;,
\end{align}
\end{subequations}
%--
where the normalization factor is given by
%--
\begin{align}\label{Nfactor}
{\cal N}_\xi = \left(\{1-(v+L)^2{d \over dz}\Sigma^+(z_\xi) \right) 
+ \left(1-(v-L)^{2}{d \over dz}\Sigma^+(z_\xi)\right)  {z_\xi+B_z-(v+L)^2 \Sigma^+(z_\xi)  \over z_\xi-B_z-(v-L)^2\Sigma^+(z_\xi)} \;.\end{align}
%--
\end{widetext}
With  use of the explicit forms of the transition matrix elements given in (\ref{AppEq:MatrixElements}), we find that each term in terms of the  states to the two-dimensional Fourier transform spectrum is proportional to the product of the normalization factors as
%--
\begin{align}\label{chicompo}
\chi_{i,j}^{(\xi,\xi',\xi'')}(\omega_2,\omega_1)\propto {1\over  {\cal N}_\xi {\cal N}_{\xi''} {\cal N}_{\xi'}^*  }  {1 \over \omega_2-\Delta_{\xi'',\xi'}} {1 \over \omega_1- \Delta_{\xi,\xi'}}  \;.
\end{align}
%--
The first factors (the normalization constants)  represent the Peterman effect, and the second and third resonance  factors (the propagators)  represent the Purcell effect\cite{Lin2016,Pick2017}.

We have also calculated the one-dimensional Fourier transform spectrum (1DFT) for the sudden excitation of the probe pulse at $t_1=0$\cite{Jonas2003}, which reads
%--
\begin{align}
&{\cal S}_{i,j}(\omega_2)
=  4 g_B  \sum_{\xi',\xi''} {\rm Re}\left[ {1 \over \omega_2-\Delta_{\xi'',\xi'}}   \right.
 \notag\\
 &\left. \times\< 0,B_+|\tilde\phi_{\xi'} \> \<\phi_{\xi'} | \hat  s_i({\bs r}_D)|\phi_{\xi''}\>  \<\tilde\phi_{\xi''} | \hat  s_j({\bs r}_D)  |0,B_+\> \right] \\
& \equiv 4 g_B  \sum_{\xi',\xi''} {\rm Re}\left[ {\cal S}_{i,j}^{(\xi',\xi'')}(\omega_2) \right] \;.
\end{align}
%-- 
Each term of  ${\cal S}_{i,j}^{(\xi',\xi'')}$ is also composed of the normalization factor and the resonance enhancement factor as
%--
\begin{align}\label{Scompo}
{\cal S}_{i,j}^{(\xi',\xi'')}(\omega_2)\propto {1\over  {\cal N}_{\xi'}^* {\cal N}_{\xi''}  } {1 \over \omega_2-\Delta_{\xi'',\xi'}}  \;.
\end{align}
%--
It is seen from  comparison of (\ref{chicompo}) and (\ref{Scompo}) that the 2DFT spectrum is more enhanced than the 1DFT spectrum.

%--------------------------------------
\begin{figure}[t]
\begin{center}
\includegraphics[width=90mm,height=80mm]{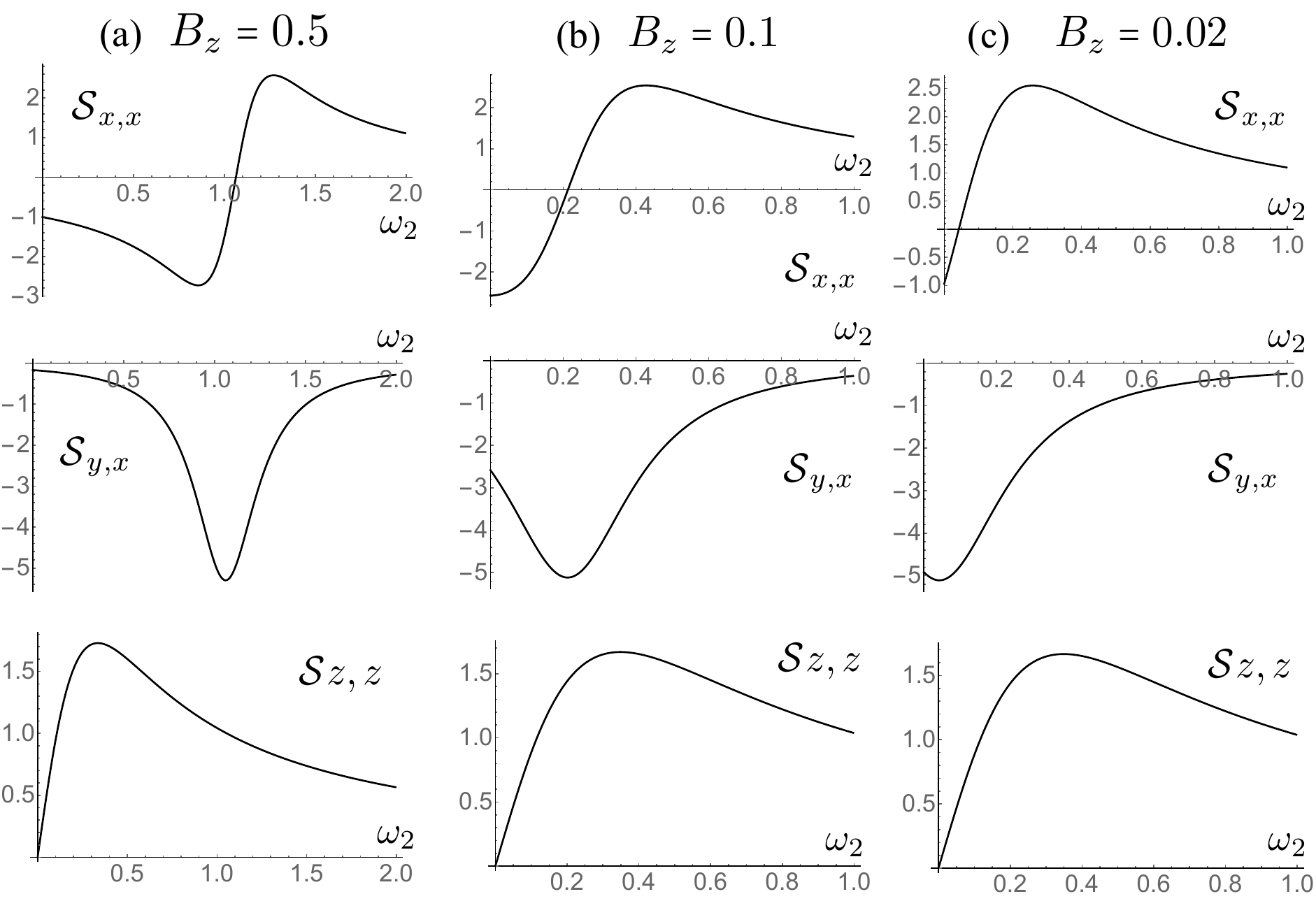}
\caption{The 1DFT SSESR spectrum ${\cal S}_{i,j}(\omega_2)/g_B$ for $B_z=0.5$ (a) and $B_z=0.1$(b), and $B_z=0.02$(c), where $B_x=B_y=0$.
The parameters are taken as $v=0.3$ and $L=0.1$.
The spectral components  ${\cal S}_{x,x}$, ${\cal S}_{y,x}$, and ${\cal S}_{z,z}$ are shown in the top, middle, and bottom rows, respectively.}
\label{fig:BZline}
\end{center}
\end{figure}
%----------------

\begin{figure*}[ht]
\begin{center}
\includegraphics[width=180mm,height=160mm]{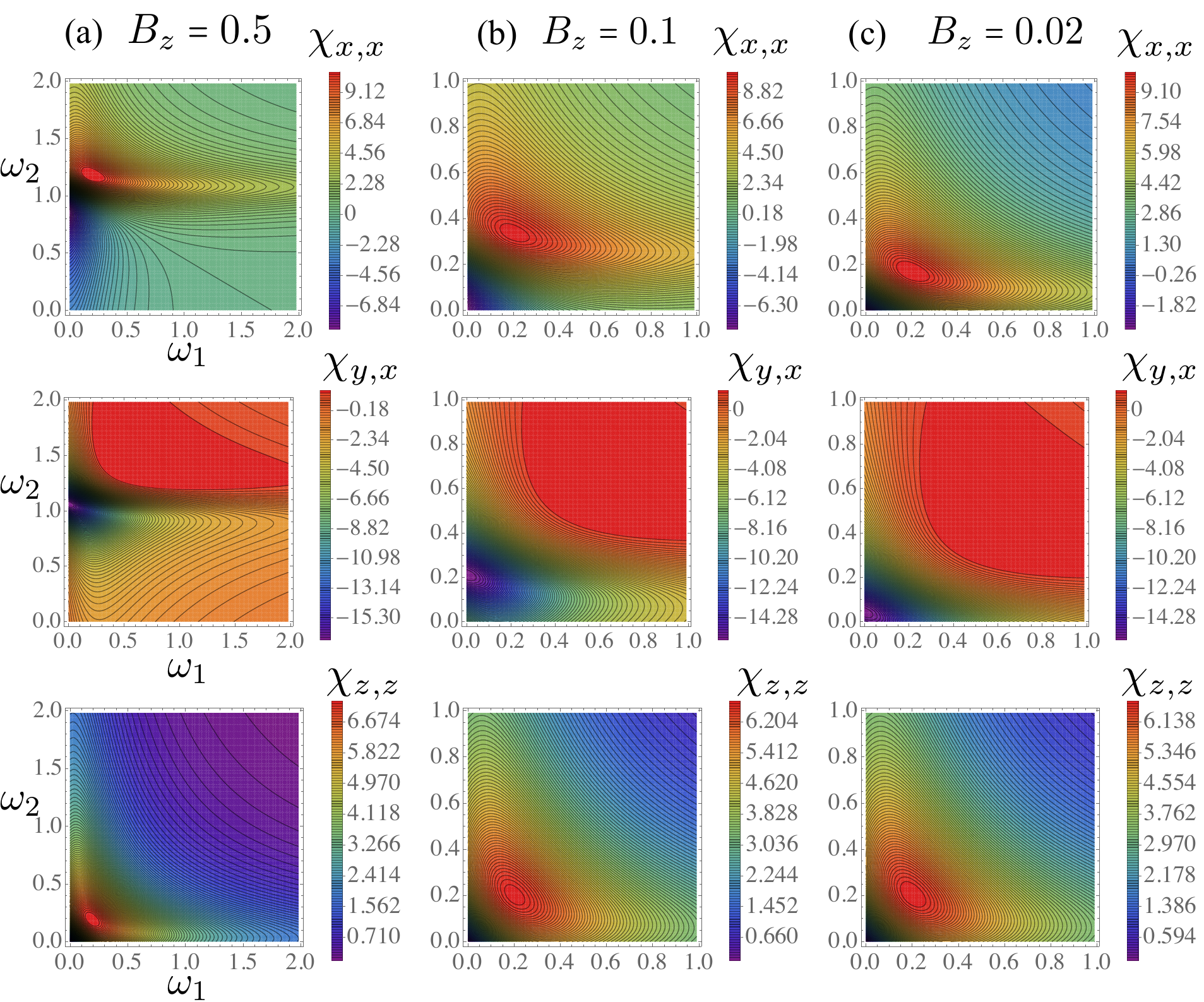}
\caption{The 2DFT SSESR spectrum ($\chi_{i,j}/g_B$) for $B_z=0.5$ (a) and $B_z=0.1$(b), and $B_z=0.02$(c), where $B_x=B_y=0$.
The parameter values are the same as in Fig.\ref{fig:BZline}.
The spectral components of $\chi_{x,x}$, $\chi_{y,x}$, and $\chi_{z,z}$ are shown in the top, middle, and bottom views, respectively.}
\label{fig:BZ1}
\end{center}
\end{figure*}

We show  in Fig.\ref{fig:BZline} the 1DFT SSESR spectrum, when the external magnetic field is directed parallel to the molecular field:  (a) $B_z=0.5$, (b) $B_z=0.1$, and (c) $B_z=0.02$, while $B_x=B_y=0$.
In this case, since the spin states are decoupled into the up- and down-spin subspaces, it is easy to evaluate the 1DFT SSESR spectrum.
Here we evaluate the ${\cal S}_{x,x}$ component shown in Fig.\ref{fig:BZline}(a) in the resonance pole approximation as 
%--
\begin{subequations}\label{BzSxx}
\begin{align}
{\cal S}_{x,x}(\omega_2)
&=  4 g_B \left| \< 0,B_+|\tilde\phi_{+} \> \right|^2 {\rm Re}\left[ {\<\phi_{+} | \hat  s_x|\phi_{-}\>  \<\tilde\phi_{-} | \hat  s_x  |\phi_{+}\>  \over \omega_2-\Delta_{\phi_-,\phi_+}}   \right] \\
&\simeq  4 g_B {\omega_2-2 B \over (\omega_2-2 B_z)^2+4 \gamma^2(v^2+L^2)^2     } \;,
\end{align}
\end{subequations}
%--
with $\gamma$ defined by
%--
\begin{align}
\gamma\equiv |{\rm Im}\Sigma(B)|={1\over 2}\sqrt{4-B^2} \;.
\end{align}
%--
In (\ref{BzSxx}a), we have used $\Delta_{\phi_-,\phi_+}\simeq 2B_z  +2i \gamma (v^2+L^2) $,  $\<\phi_{+} | \hat  s_x|\phi_{-}\> \simeq \<\tilde\phi_{-} | \hat  s_x  |\phi_{+}\> \simeq 1$, and $\left| \< 0,B_+|\tilde\phi_{+} \> \right|^2\simeq 1$, as evaluated in \eqref{AppEq:MatrixElements}, which are good approximations for $B_z\ll 2$.
Therefore, the ${\cal S}_{x,x}$ component is represented by a dispersion-type function with the node at $\omega_2=2B$ and the width of $2\gamma (v^2+L^2)$, as shown in the top rows of  Fig.\ref{fig:BZline}.
Similarly, the ${\cal S}_{y,x}$ component is given by the Lorentz function with the peak at $\omega_2=2B$ and the width   $2\gamma (v^2+L^2)$ as
%--
\begin{subequations}\label{BzSyx}
\begin{align}
{\cal S}_{y,x}(\omega_2)
&=4g_B\left| \<0,B_+|\tilde\phi_+\>\right|^2 {\rm Re}\left[  { \<\phi_{+} | \hat  s_y|\phi_{-}\>  \<\phi_{-} | \hat  s_x  | \phi_+\>   \over \omega_2-\Delta_{\phi_-,\phi_+}}  \right] \\
&\simeq  -8 g_B  { 2\gamma (v^2+L^2)  \over (\omega_2-2 B)^2+4 \gamma^2(v^2+L^2)^2     } \;,
\end{align}
\end{subequations}
%--
and the ${\cal S}_{z,z}$ component is given by the dispersion-type function with the node at $\omega_2=0$ and the width of $2\gamma (v+L)^2$ as
%--
\begin{subequations}\label{BzSzz}
\begin{align}
{\cal S}_{z,z}(\omega_2)
&=  4 g_B   \left| \< 0,B_+|\tilde\phi_{+} \> \right|^2 {\rm Re}\left[ { \<\phi_{+} | \hat  s_z|\phi_{+}\>  \<\phi_{+} | \hat  s_z  |\phi_+\>  \over \omega_2-\Delta_{\phi_+,\phi_+}}   \right] \\
&\simeq  4 g_B  { \omega_2  \over \omega_2^2+4 \gamma^2(v+L)^4     } \;.
\end{align}
\end{subequations}
%--
as shown in the middle and the bottom rows of Fig.\ref{fig:BZline}, where we have used $\<\phi_{+} | \hat  s_y|\phi_{-}\> \simeq i$ in (\ref{BzSyx}a), and  $\<\phi_{+} | \hat  s_z|\phi_{+}\> \simeq 1$ and  $\Delta_{\phi_+,\phi_+}\simeq  -2i \gamma (v+ L)^2$ in (\ref{BzSzz}a).
Due to the symmetry, the other components of ${\cal S}_{x,z}$, ${\cal S}_{y,z}$, ${\cal S}_{z,x}$, and ${\cal S}_{y,z}$ vanish.
Comparing these results with the complex eigenvalues in Fig.\ref{fig:E0}, we find that the spectrum clearly reflects the features of  the complex eigenvalues.
We show the 2DFT SSESR spectrum in Fig.\ref{fig:BZ1} for the same conditions. 
The cross correlation between $\omega_1$ and $\omega_2$ provides us with  detailed information about the spin relaxation at the impurity site.

%--------------------------------------
%----------------

%--------------------------------------
\begin{figure*}[htp]
\begin{center}
\includegraphics[width=180mm,height=220mm]{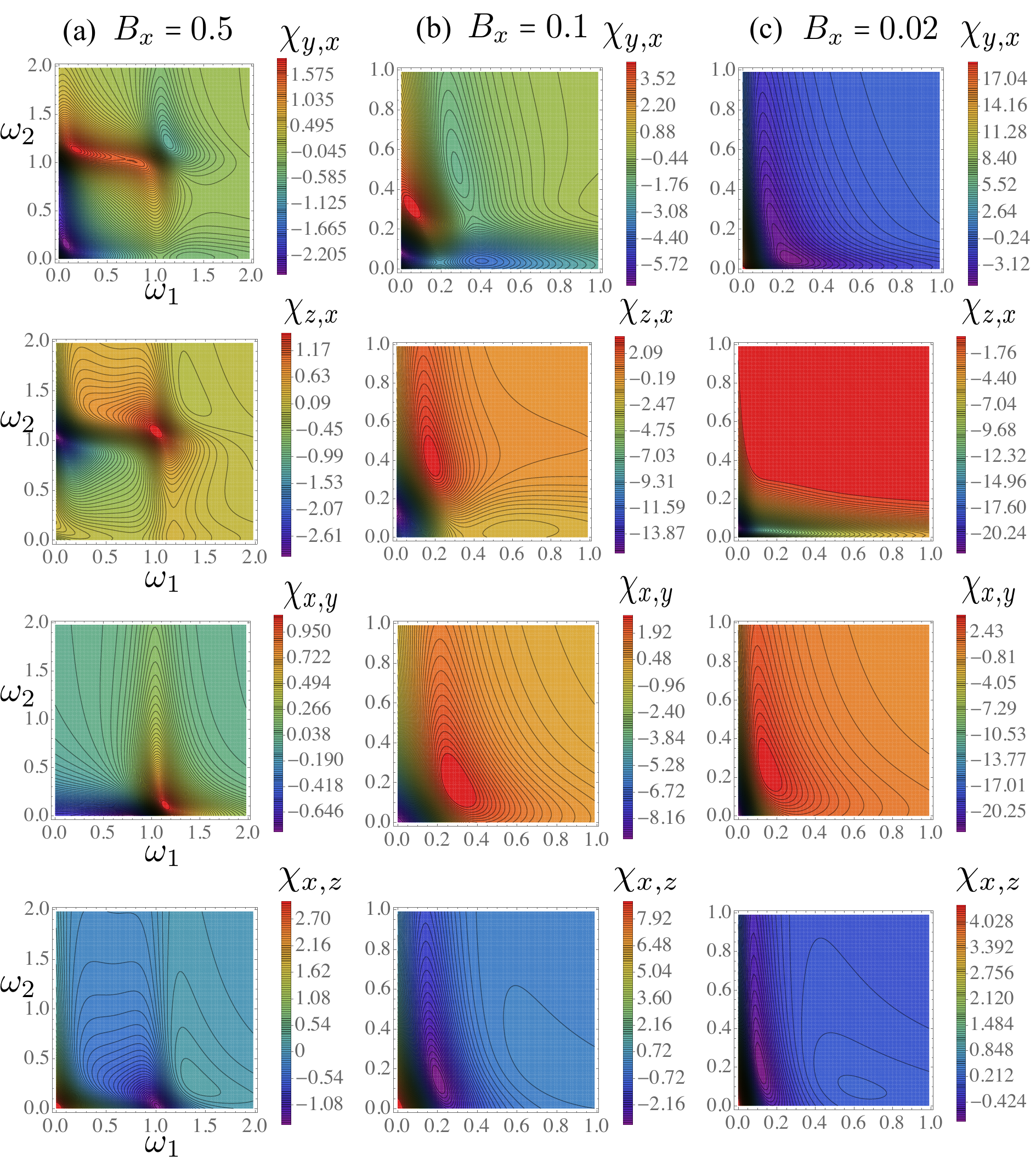}
\caption{The 2DFT SSESR spectrum ($\chi_{i,j}/g_B$)  for  $B_x=0.5$ (a) and $B_x=0.1$(b), and $B_x=0.02$(c), where $B_z=B_y=0$ for the same parameters as in Fig.\ref{fig:EPring}.
The first, second,  third, and  fourth rows are $\chi_{y,x}$, $\chi_{z,x}$, $\chi_{x,y}$, and $\chi_{x,z}$ components, respectively, which are only allowed by the introduction of the spin-orbit coupling. }
\label{fig:BX52}
\end{center}
\end{figure*}
%----------------------------------
%
%%--------------------------------------
%\begin{figure}[t]
%\begin{center}
%\includegraphics[width=90mm,height=60mm]{BX52line}
%\caption{The 1DFT SSESR spectrum ${\cal S}_{i,j}(\omega_2)$ for $B_x=0.5$ (a) and $B_x=0.1$(b), and $B_x=0.02$(c), where $B_x=B_y=0$, corresponding to Fig.\ref{fig:BX52}.
%The spectral components of ${\cal S}_{y,x}$ and ${\cal S}_{z,x}$ are shown in the top and bottom views, respectively.}
%\label{fig:BXline}
%\end{center}
%\end{figure}
%%----------------

As shown in Figs.\ref{fig:EPmiddle} and \ref{fig:EPring}, the spin states are coupled when the external magnetic field is tilted away from the molecular field whose direction is governed by the molecular axis through the spin-orbit interaction.
The change of the spin states is reflected in the 2DFT SSESR as an anisotropy in terms of the relative angle between the probe field and the detection direction.
We show in Fig.\ref{fig:BX52} the 2DFT SSESR spectrum when the external magnetic field is perpendicular to the molecular field:  (a) $B_x=0.5$, (b) $B_x=0.1$, and (c) $B_x=0.02$, for  $B_y=B_z=0$, where the first, second,  third, and  fourth rows are $\chi_{y,x}$, $\chi_{z,x}$, $\chi_{x,y}$, and $\chi_{x,z}$ components, respectively.
This figure corresponds to the case of Fig.\ref{fig:EPring}. 
Although these spectral components are forbidden by the selection rule for the far-off-resonant case (\ref{Selection}), they become allowed by the introduction of the spin-orbit coupling whose effect becomes prominent by the resonance factor.
The cross correlation between $\omega_1$ and $\omega_2$ provides with the detailed information on the spin relaxation process at a donor site.

%--------------------------------------
\begin{figure}[htp]
\begin{center}
\includegraphics[width=85mm,height=80mm]{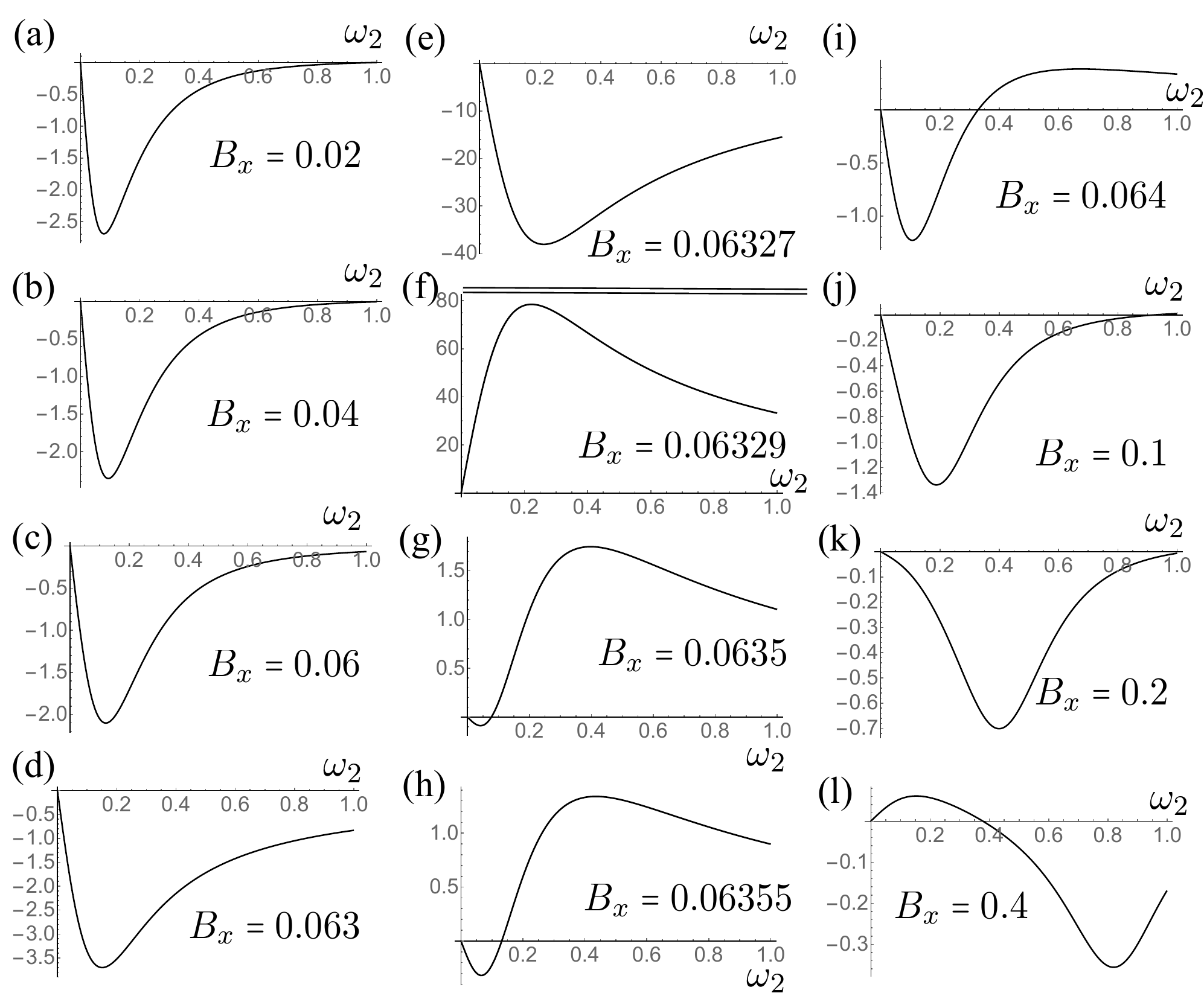}
\caption{The ${\cal S}_{z,x}/g_B$ component of the 1DFT SSESR around the EP ring singularity, which occurs at $B_x=B_{\rm EP,-}=0.0632808$. 
The double line between (e) and (f) indicates a border of the EP ring.}
\label{fig:BxSzx}
\end{center}
\end{figure}
%----------------------------------

Lastly, we show the most striking feature due to the presence of the EP ring, {\it the giant response} of the single-spin resonance.
For the illustration, we show in Fig.\ref{fig:BxSzx} the ${\cal S}_{z,x}$ component the 1DFT SSESR for various values of $B_x$ across the EP ring singularity, corresponding to Fig.\ref{fig:EPring}(c) and (d).
%This spectral component is composed of the two contribution: the Zeeman split resonance states, and the self-state. 
As the value of $B_x$ decreases from (l) to (i), the negative peak position shifts to the lower frequency side.
The peak position reflects the real Zeeman splitting as $2B_x$, as shown in Fig.\ref{fig:EPring}(c).
As $B_x$ further decreases and comes close to the  $B_{\rm EP,-}$ singularity, the signal amplitude becomes positively large, as seen in (f).
When $B_x$ crosses over $B_{\rm EP,-}$, the sign of the signal is flipped keeping the large signal amplitude, as shown in (e).
Then as $B_x$ decreases, the signal amplitude becomes small with the dispersion-type function with a node at $B_x$=0, as shown in (d) to (a).

%--------------------------------------
\begin{figure*}[htp]
\begin{center}
\includegraphics[width=180mm,height=170mm]{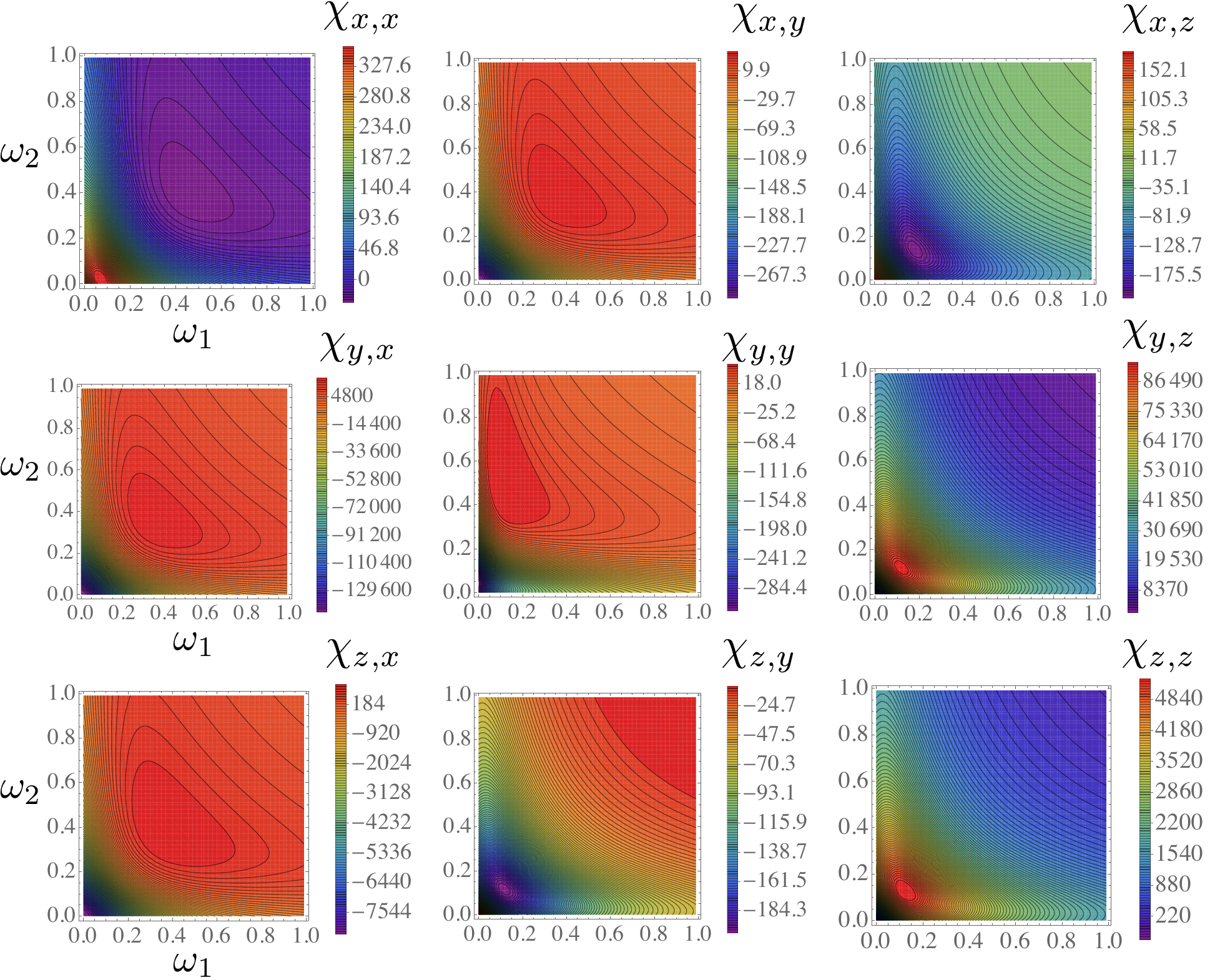}
\caption{The 2DFT SSESR  spectrum  $(\chi_{i,j}/g_B)$ at $B_x=0.06327\simeq B_{{\rm EP},-}$.}
\label{fig:BxRing}
\end{center}
\end{figure*}
%----------------------------------

The giant response at the EP ring singularity is much more pronounced in the 2DFT SSESR spectrum.
We show in Fig.\ref{fig:BxRing} the 2DFT SSESR spectral components at $B_x=0.06327$ corresponding to Fig.\ref{fig:BxSzx}(e).
We have found the extremely huge signal amplitude in the spectrum, because of the enhancement due to the normalization constant and the resonance factor as mentioned in (\ref{chicompo}).
This giant response of the SSESR at the EP singularity may be very useful to observe the single-spin resonance which is very weak in the ordinary situation.
Therefore, this method becomes powerful tool to observe the spin relaxation process with an atomic scale precision.

%##################################
\section{Concluding Remarks}\label{Sec:Conclusion}

We have studied the complex eigenenergy structure associated with the charge transfer decay of an alkali-doped polyacetylene under an external static magnetic field.
Starting with the hermitian Hamiltonian of the total system, we have derived the non-hermitian effective Hamiltonian, where the microscopic spin-dependent decay processes are incorporated in terms of the energy-dependent self-energy.
The dynamical and chiral symmetries that the total Hamiltonian possesses are maintained in the effective Hamiltonian.

We have found that the spin-orbit interaction influences the Zeeman splitting of the spin states at the donor site in the charge transfer decay.
As a result, the complex eigenenergy structure exhibits a strong anisotropy in the external magnetic field: The EP singularity due to the chiral symmetry breaking appears when the magnetic field is perpendicular to the molecular field to form the EP ring. In contrast, the EP singularity due to the dynamical symmetry breaking appears irrespective of the field direction yielding the EP surface.

We have revealed that the coherent single-spin resonance spectrum is a powerful tool to observe the complex eigenenergy spectrum of the system with atomic precision by using its sensitivity to the directions of the probe pulse and the detection.
For the formulation of the SSESR, we have utilized the nonlinear response function in the Liouville-space pathway approach, where we have constructed the Liouville space basis in terms of the complex eigenstates of the total Hamiltonian.
We  emphasize the importance of the energy dependence of the self-energy in the Liouville-space formalism.
In the Liouville-space formalism, we need pairs of the resonance and anti-resonance states with complex conjugate eigenvalue pairs of $z$ and $z^*$ so that the Liouville basis be consistent with the second law of thermodynamics, i.e., entropy production\cite{Petrosky91Physica,Ordonez01PRA,Ordoez2001}.
The present method properly deals with the dynamical symmetry of the total hermitian Hamiltonian in the effective Hamiltonian by taking into account the energy dependence of the self-energy.

We have demonstrated that the 1DFT and 2DFT SSESR spectra probe the spin-relaxation dynamics at the donor site.
While the 1DFT SSESR spectrum reflects the complex eigenenergy spectrum, the 2DFT gives detailed information on the quantum coherence in the spin-relaxation dynamics as a cross-correlation between the two frequencies.
We found a giant response of the coherent SSESR around the EP ring singularity due to the vanishing normalization factors at the EP ring and the resonance effect.
We have discovered that the giant response is much larger in magnitudes in the 2DFT spectrum than in the 1DFT spectrum.
Therefore, the 2DFT SSESR spectroscopy becomes a promising tool to observe the single-spin response in a molecule in contrast to the ordinary linear response, such as absorption or spontaneous emission spectroscopies\cite{Lin2016,Pick2017a}.

There are several issues worth to be further explored.
In this work, we have focused on the spin-polarized charge transfer decay in a polyacetylene due to the spin-orbit coupling.
The charge transfer in polyacetylene has been investigated in terms of the soliton or polaron formation accompanied by the structural deformation.
Although we do not take into account the dynamical change of the molecular structure in this work because it is considered to be much slower than the electronic motion,
it is worth  studying the influence of the soliton or polaron motion on the single-spin resonance spectroscopy.

We have revealed in Section \ref{Sec:EVHee} the appearance of the two EP singularities: EP surface and EP ring.
If the spin-orbit interaction is large or conduction bandwidth is small, these two singularities will merge. 
It is interesting to study the response at that merged singularities, where the non-Markovian effect is expected to be largely enhanced\cite{Garmon2017a,Garmon2019} 

For the detailed study very close to the exceptional point, it would be better to use pseudo-eigenstate representations\cite{kato2012short}.
Otherwise the resonant and anti-resonant state representations may increase numerical error very close to the EP singularity, because the coalesced state at an EP is self-orthogonal and cannot be normalized.
Recently, Hashimoto and Kanki {\it et al.} have invented {\it extended Jordan block basis}  which is continuously connected to the Jordan block exactly at the exceptional point.
The extended Jordan block basis provides a useful way to describe the observable quantities continuously across the exceptional point\cite{Kanki16Springer,Hashimoto2016,Hashimoto2016a,Kanki2017}.
This representation may remove some difficulties in calculating the ESR spectrum close to the EP singularity.
Moreover, we have an exceptional point of order 4 (EP4) of the Liouvillian, where four eigenstates coalesce, when the eigenstate of the Liouvillian is given in the form \eqref{AppEq:Lbasis} and both of the eigenstates of the Hamiltonian in \eqref{AppEq:Lbasis} tend to the coalesced states at an EP of the Hamiltonian.
As a result, the Liouvillian contains a $4\times 4$ Jordan block, and the resolvent of the Liouvillian has a fourth order pole.
This fourth order pole structure may be reflected in Fig.\ref{fig:BxRing}.
Detailed analysis of this EP4 will be performed in the future.

In the present work, we have adopted the impulsive pulse limit in the SSESR spectrum to clarify the effect of the EP singularity.
Meanwhile, the recent advancement in the pulse control technique with the use of phase-locked laser pulse, such as the carrier-envelope-phase (CEP) control\cite{Sell2008}, enables us to clarify the single-spin relaxation process in more detail, such as quantum decoherence as well as population decay.

For the SSESR calculation, we have used the time-dependent perturbation method in terms of the probe magnetic pulse field, assuming a weak probe field intensity.
However, this treatment is to be reexamined when close to the EP singularity, because the interaction energy with the probe pulse becomes very large by the giant response even though the field intensity is weak.
For the strong time dependent external filed, the Floquet method may be applied to the present system in the Liouville space\cite{Oka19AnnRev,Grifoni1998,sym10080313}.

%\clearpage

%\bibliography{/Users/SatoshiTanaka/Dropbox/Physics/Presentation/BibTeX/library} 

\acknowledgments

We would like to thank T. Petrosky, Y. Kayanuma, K.-i. Noba, K. Hashimoto, and K. Imura for fruitful discussions.
This research was funded  by JSPS KAKENHI grants number  JP18K03496 and JP17K05585.

\appendix
%###############
\section{Matrix elements of the orbital angular momentum between the atomic orbitals}\label{AppSec:SO}

In this section, we evaluate the matrix elements of the orbital angular momentum between the $ns$ orbital of the heavy alkali donor atom and the $2p_y$ orbital of the carbon atom at the end of the molecule.
The matrix elements are given by
%--
\begin{align} \label{AppEq:lielement}
l_{i}\equiv \<0,s|\hat{l}_{i}|1,p_y\>=\int d^{3}\boldsymbol{r}\varphi_{s}^{*}(\boldsymbol{r})\hat{l}_{i}\varphi_{y}(\boldsymbol{r}-\boldsymbol{a})
\end{align}
%--
where we locate the $ns$ donor atomic orbital $\varphi_{z}$ at the origin, and the $2p_y$ carbon atomic orbital $\varphi_{y}$ is centered at ${\boldsymbol r}={\boldsymbol a}$, as shown in Fig.\ref{fig:Molecule}.
The wave functions are given by
%-
\begin{align} \label{AppEq:pz}
&\varphi_{y}(\mathbf{r})=R_{21}(r)\{Y_{1,1}(\theta,\phi)+Y_{1,-1}(\theta,\phi)\} \;,\\
 &Y_{1,1}(\theta,\phi)+Y_{1,-1}(\theta,\phi)=\sqrt{\frac{3}{8\pi}}\sin\theta\sin\phi \;,
\end{align}
%--
and the alkali atom $ns$-orbital as
%-
\begin{align} \label{AppEq:s}
\varphi_s({\bf r})=\varphi_s( r)={1\over \sqrt{4\pi}} R_{ns}(r) e^{-r} \;.
\end{align}
%--
Substituting (\ref{AppEq:pz}) into (\ref{AppEq:lielement}), we have after simple calculation
%--
\begin{align} \label{AppEq:lz1}
l_{z} & =\sqrt{\frac{3}{4\pi}}\int d^{3}\boldsymbol{r}\varphi_{s}^{*}(\boldsymbol{r})e^{-ia\hat{p}_{x}}  \frac{1}{i}(a\sin^{2}\theta\sin^{2}\phi\frac{\partial}{\partial r}R_{21}(r) \notag\\&+a\frac{\cos^{2}\theta\sin^{2}\phi-\cos^{2}\phi}{r}R_{21}(r)-R_{21}(r)\sin\theta\cos\phi)
\end{align}
%--
Similarly we obtain for $\hat l_x$ and $\hat l_y$ elements as
%--
\begin{align}
l_{x} &=\sqrt{\frac{3}{\pi}}\int d^{3}\boldsymbol{r}\varphi_{s}^{*}(\boldsymbol{r}+\boldsymbol{a})R_{21}(r)\cos\theta  \label{AppEq:lx1} \\
l_{y} & =\sqrt{\frac{3}{4\pi}}a\int d^{3}\boldsymbol{r}\varphi_{s}^{*}(\boldsymbol{r}+\boldsymbol{a})\frac{1}{i}\sin\theta\cos\theta\sin\phi \notag\\
&\times (\frac{\partial}{\partial r}R_{21}(r)-\frac{1}{r}R_{21}(r)) \;.  \label{AppEq:ly1}
\end{align}
%--

Substituting (\ref{AppEq:s}) into (\ref{AppEq:lz1}), (\ref{AppEq:lx1}), and (\ref{AppEq:ly1}), and executing the solid angle integral, we find
%--
\begin{align}
l_{z}=\frac{3}{16}\sqrt{\frac{3}{2}}\pi a\int drr^{2}R_{ns}(r+a)e^{-(r+a)}\frac{1}{i}\frac{\partial}{\partial r}R_{21}(r)
\end{align}
while
%--
\begin{align}
l_x=l_y=0 \;.
\end{align}
%--
Radial integral can be performed once the radial wave function of the atomic orbitals are given.

%##################################
\section{Chiral symmetry of the total Hamiltonian}
\label{AppSec:GeoHtot}

In this section we show the chiral symmetry of the total system Hamiltonian in the $\bf B$-parameter space.
For this purpose, we represent the total Hamiltonian in terms of the alternate site basis which is aligned as
%--
\begin{align}
\{ |0,+\> ,  |1,-\> ,  |2,+\> , \cdots ,  |0,-\> ,  |1,+\>  , |2,-\> \} \;.
\end{align}
%--
The total Hamiltonian is represented by
%--
\begin{widetext}
%--
\begin{align}\label{AppEq:LFfull}
\hat H_{\rm tot}= 
\begin{array}{c| c c c c c  c c c }
 &  |0,+\> &  |1,-\>  &  |2,+\>  & \cdots &  |0,-\> &  |1,+\>  & |2,-\> & \cdots  \\ \hline
\< 0,+| & \varepsilon_D-B_z   & 0  &0  & \cdots & -B_x + iB_y  &  v+L  & 0 &\cdots\\
\< 1,-| & 0 & 0  &0  & \cdots & v-L &  0 & -J &\cdots\\
\< 2,+| & 0 & 0  &0  & \cdots & 0  &  -J & 0 &\cdots\\
\vdots & \vdots & \vdots  & \vdots & \vdots & \vdots & \vdots & \vdots & \vdots \\
\< 0,-| & -B_x-iB_y & v-L  &0  & \cdots & \varepsilon_D+B_z  & 0  & 0 &\cdots\\
\< 1,+| & v+L & 0  &-J  & \cdots & 0 &  0 & 0&\cdots\\
\< 2,-| & 0 & -J  &0  & \cdots & 0  &  0 & 0 &\cdots\\\end{array} \;.
\end{align}
\end{widetext}
%--
For $\varepsilon_D=0$, we see that the total Hamiltonian satisfies a chiral symmetry \cite{Ryu2010} when $B_z=0$
%--
\begin{align}\label{AppEq:HtotSym}
\boldsymbol{\sigma}_z \hat H_{\rm tot} \boldsymbol\sigma_z =-\hat H_{\rm tot} \;,
\end{align}
%--
where ${\boldsymbol\sigma}_z$ is an infinite dimensional matrix for alternate site basis
%--
\begin{align}
\boldsymbol\sigma_z \equiv \begin{pmatrix}
I & 0\\
0 & -I
\end{pmatrix} \;.
\end{align}
%--

It follows from \eqref{AppEq:HtotSym} that
%--
\begin{subequations}\label{AppEq:HtotSymEV}
\begin{align}
&\hat H_{\rm tot}|\psi\>=z|\psi\>  \;,\\
\implies &\boldsymbol\sigma_z\hat H_{\rm tot}\bar{\boldsymbol\sigma}_z \bar{\boldsymbol\sigma}_z |\psi\>=z\boldsymbol\sigma_z|\psi\>  \;,\\
\implies &\hat H_{\rm tot}  \bar{\boldsymbol\sigma}_z |\psi\>=-z  \bar{\boldsymbol\sigma}_z |\psi\>  \;,
\end{align}
\end{subequations}
%--
which suggests that the eigenvalues of the total Hamiltonian are obtained as a pair of $z$ and $-z$ at $B_z=0$.
We notice that when $\varepsilon_D\neq 0$ or $B_z\neq 0$, the chiral symmetry is not satisfied, so that the EP ring does not appear.
We have shown that  the total Hamiltonian has a property that it exhibits the chiral symmetry at the special chiral configuration in the $\bf B$-space, i.e., at $B_z=0$.

%##################################
\section{Complex eigenvalue problem of the total Hamiltonian}
\label{AppSec:Heff}

In this section, we derive the effective Hamiltonian in terms of the BWF projection method.
We consider the projection operator onto the two-spinor at the donor site given by
\begin{equation}
\hat{P}_0\equiv \sum_{\alpha=\pm}|0, \alpha\>\<0,\alpha| \;,
\end{equation}
%--
and its complement
%--
\begin{equation}
\hat{Q}_0\equiv1-\hat{P}_0= \sum_{\alpha=\pm} \int dk|k,\alpha\>\<k,\alpha| \;.
\end{equation}
%--
Acting these projection operators on (\ref{HtotEVP}) from the left, we have
%--
\begin{subequations}
\begin{align}
\hat{P}_{0}\hat{H}_{tot}\hat{P}_{0}|\phi_{\xi}\>+\hat{P}_{0}\hat{H}_{tot}\hat{Q}_{0}|\phi_{\xi}\> & =z_{\xi}\hat{P}_{0}|\phi_{\xi}\>\;,\\
\hat{Q}_{0}\hat{H}_{tot}\hat{P}_{0}|\phi_{\xi}\>+\hat{Q}_{0}\hat{H}_{tot}\hat{Q}_{0}|\phi_{\xi}\> & =z_{\xi}\hat{Q}_{0}|\phi_{\xi}\>\;.
\end{align}
\end{subequations}
%--

The effective Hamiltonian is given by
%--
\begin{align}\label{AppEq:HeffForm}
\hat H_{\rm eff}({\bf B};z)=\hat P_0 \hat H_{\rm tot} \hat P_0+\hat P_0 \hat H_{\rm tot} \hat Q_0{1\over z-\hat Q_0 \hat H_{\rm tot} \hat Q_0}\hat Q_0 \hat H_{\rm tot} \hat P_0 \;.
\end{align}
%--
Evaluating this for the present model gives Eq.(\ref{Heff}).

Then the complex eigenvalue problem of the effective Hamiltonian reads
%--
\begin{align}
\hat H_{\rm eff}({\bf B};z) \hat P_0|\phi_\xi\>=z_\xi \hat P_0|\phi_\xi\> \;,\; \<\tilde\phi_\xi|\hat P_0 \hat H_{\rm eff}({\bf B};z) =z_\xi  \<\tilde\phi_\xi|\hat P_0\;.
\end{align}
%--
The explicit form of the eigenvalue problem then reads
%--
\begin{align}\label{AppEq:HeffEVR}
&\begin{pmatrix}
-B_{z}+(v+L)^{2}\Sigma^+(z_\xi) &- B_{x}+iB_{y}\\
-B_{x}-iB_{y} & B_{z}+(v-L)^{2}\Sigma^+(z_\xi)
\end{pmatrix}
\begin{pmatrix}
\<0,+|\phi_\xi\>\\
\<0,-|\phi_\xi\>
\end{pmatrix} \notag\\
& \hspace{2cm}
=z_\xi
\begin{pmatrix}
\<0,+|\phi_\xi\>\\
\<0,-|\phi_\xi\>
\end{pmatrix}
\end{align}
%--
for the right-eigenstates, and 
%--
\begin{align}\label{AppEq:HeffEVL}
&
\begin{pmatrix}
\<\tilde\phi_\xi|0,+\>\\
\<\tilde\phi_\xi|0,-\>
\end{pmatrix}^T
\begin{pmatrix}
-B_{z}+(v+L)^{2}\Sigma^+(z_\xi) &- B_{x}+iB_{y}\\
-B_{x}-iB_{y} & B_{z}+(v-L)^{2}\Sigma^+(z_\xi)
\end{pmatrix} \notag\\
&\hspace{2cm}
=z_\xi
\begin{pmatrix}
\<\tilde\phi_\xi|0,+\>\\
\<\tilde\phi_\xi|0,-\>
\end{pmatrix}^T
\end{align}
%--
for the left-eigenstates.
From these eigenvalue problems, we see the relations
%--
\begin{subequations}
\begin{align}
{\<0,-|\phi_\xi\>\over \<0,+|\phi_\xi\>}&={B_x+ i B_y \over  B_z+(v-L)^2 \Sigma^+(z_\xi)-z_\xi }  \;,\\
{\<\tilde\phi_\xi|0,-\>\over\<\tilde\phi_\xi|0,+\>}&={B_x- i B_y \over  B_z+(v-L)^2 \Sigma^+(z_\xi)-z_\xi} \;,
\end{align}
\end{subequations}
%--
which leads to
%--
\begin{align}\label{AppEq:phiphi}
{\<\tilde\phi_\xi|0,-\>\<0,-|\phi_\xi\>\over \<\tilde\phi_\xi|0,+\>\<0,+|\phi_\xi\>}={B_x^2 + B_y^2 \over ( B_z+(v-L)^2 \Sigma^+(z_\xi)-z_\xi)^2 } \;.
\end{align}
%--

The eigenstates of the total Hamiltonian are obtained by adding the $\hat Q_0$-component as
%--
\begin{align}
|\phi_\xi\>&=\hat P_0|\phi_\xi\>+ \hat Q_0|\phi_\xi\> \notag\\
&=\hat P_0|\phi_\xi\>+{1\over z_\xi-\hat Q_0 \hat H_{\rm tot} \hat Q_0}\hat Q_0 \hat H_{\rm tot}\hat P_0|\phi_\xi\>
\end{align}
%--
In the present system, we have obtained
%--
\begin{align}\label{AppEq:phixi}
&|\phi_\xi\>= \<0,+|\phi_\xi \> \left\{ |0,+\> + (v+L) \sqrt{2\over \pi}\int_0^\pi dk {\sin k \over (z-\omega_k)_{z_\xi}^+}|k,+\> \right\} \notag\\
&+  \<0,-|\phi_\xi \> \left\{ |0,-\> + (v-L) \sqrt{2\over \pi}\int_0^\pi dk {\sin k \over (z-\omega_k)_{z_\xi}^+}|k,-\> \right\} 
\end{align}
%--
for the discrete right-eigenstate with the complex eigenvalue of $z_\xi$.
Similary the discrete left eigenstates are obtained as
%--
\begin{align}\label{AppEq:leftphi}
&\<\tilde\phi_\xi| = \<\tilde\phi_\xi|0,+ \> \left\{ \<0,+ | + (v+L) \sqrt{2\over \pi}\int_0^\pi dk {\sin k \over (z-\omega_k)_{z_\xi}^+}\<k,+| \right\} \notag\\
&+  \<\tilde\phi_\xi|0,- \> \left\{ \<0,-| + (v-L) \sqrt{2\over \pi}\int_0^\pi dk {\sin k \over (z-\omega_k)_{z_\xi}^+}\<k,-| \right\} \;.
\end{align}
%--
The normalization factors are determined by the bi-orthonormality between the right- and left-eigenstates as
%--
\begin{align}
\<\tilde\phi_\xi|\phi_{\xi'}\>=\delta_{\xi,\xi'} \;.
\end{align}
%--
The normalization condition requires
%--
\begin{align}\label{AppEq:OrthoNorm}
1&=\<\tilde\phi_\xi| \phi_\xi\> \notag\\
&= \<\tilde\phi_\xi|0,+ \> \<0,+|\phi_\xi\> \left\{ \left( 1-(v+L)^2{d\over dz}\Sigma^+(z)\Big|_{z=z_\xi}  \right) \right. \notag\\
&\quad \left. + {  \<\tilde\phi_\xi|0,- \> \<0,-|\phi_\xi\> \over  \<\tilde\phi_\xi|0,+ \> \<0,+|\phi_\xi\>}
   \left( 1-(v-L)^2{d\over dz}\Sigma^+(z)\Big|_{z=z_\xi} \right) \right\}  \;.
\end{align} 
%--   

\begin{widetext}

Substituting (\ref{AppEq:phiphi}) into (\ref{AppEq:OrthoNorm}), we obtain
 the normalization factor as
%--
\begin{subequations}\label{AppEq:Nfactor}
\begin{align}
{\cal N}_\xi \equiv \left( \<\tilde\phi_\xi|0,+ \>\<0,+|\phi_\xi\>\right)^{-1} &= \left(\{1-(v+L)^2{d \over dz}\Sigma^+(z_\xi) \right)
 + \left(1-(v-L)^{2}{d \over dz}\Sigma^+(z_\xi)\right) { (B_x^2+B_y^2)  \over \left( B_z+(v-L)^2\Sigma^+(z_\xi)-z_\xi \right)^2} \\
&= \left(\{1-(v+L)^2{d \over dz}\Sigma^+(z_\xi) \right) 
+ \left(1-(v-L)^{2}{d \over dz}\Sigma^+(z_\xi)\right)  {z_\xi+B_z-(v+L)^2 \Sigma^+(z_\xi)  \over z_\xi-B_z-(v-L)^2\Sigma^+(z_\xi)} \;.
\end{align}
\end{subequations}
%---
which is the same as (\ref{Nfactor}).
Taking 
%--
\begin{align}
\<\tilde\phi_\xi|0,+ \>= \<0,+|\phi_\xi\>={\cal N}_\xi^{-1/2}  \;,
\end{align}
%--
the right- and left-eigenstates for the resonance states are represented by
%----
\begin{subequations}\label{AppEq:EigenStates}
\begin{align}
|\phi_\xi\>&={1\over {\cal N}_\xi^{1/2} }\left[  \left\{ |0,+\> + (v+L) \sqrt{2\over \pi}\int_0^\pi dk {\sin k \over (z-\omega_k)_{z_\xi}^+}|k,+\> \right\} \right. \notag\\
&\left. + {B_x+ i B_y \over  B_z+(v-L)^2 \Sigma^+(z_\xi)-z_\xi }  \left\{ |0,-\> + (v-L) \sqrt{2\over \pi}\int_0^\pi dk {\sin k \over (z-\omega_k)_{z_\xi}^+}|k,-\> \right\}  \right]    \;,\\
\<\tilde\phi_\xi| &={1\over {\cal N}_\xi^{1/2} } \left[  \left\{ \<0,+ | + (v+L) \sqrt{2\over \pi}\int_0^\pi dk {\sin k \over (z-\omega_k)_{z_\xi}^+}\<k,+| \right\} \right. \notag\\
&\left. +{B_x- i B_y \over  B_z+(v-L)^2 \Sigma^+(z_\xi)-z_\xi} \left\{ \<0,-| + (v-L) \sqrt{2\over \pi}\int_0^\pi dk {\sin k \over (z-\omega_k)_{z_\xi}^+}\<k,-| \right\} \right] \;.
\end{align}
\end{subequations}
%--

It should be noted that since the exceptional point is given by the simultaneous solution of the equations
%--
\begin{align}
\eta^+(z_\xi)=0 \;, {d\over dz}\eta^+(z)=0 \;,
\end{align}
%--
where $\eta^+(z)=0$ is the characteristic equation given in (\ref{CharaEq}), ${\cal N}_\xi=0$ at the exceptional point.
This indicates that the we cannot normalize the two independent resonance states at the exceptional point \cite{kato2012short,Hashimoto2016,Kanki16Springer,Kanki2017}.

With use of the explicit form of the eigenstates of the total Hamiltonian, the transition matrix elements in the spectrum are evaluated as
%--
\begin{subequations}\label{AppEq:MatrixElements}
\begin{align}
\<\tilde\phi_\xi|0,B_+\> &={1\over {\cal N}_\xi^{1/2} }   \left\{ \<0,+ |0,B_+\>+\<0,-|0,B_+\> {B_x- i B_y \over  B_z+(v-L)^2 \Sigma^+(z_\xi)-z_\xi} \right\}    \;, \\
\<\tilde\phi_{\xi''}|\hat s_j|\phi_\xi\>&={1\over {\cal N}_{\xi''}^{1/2} }{1\over {\cal N}_\xi^{1/2} }  \Bigg\{ \<0,+|\hat s_j|0,+\>  \notag  \\
&+  \<0,+|\hat s_j|0,-\>  {B_x+ i B_y \over  B_z+(v-L)^2 \Sigma^+(z_\xi)-z_\xi }   \notag \\
&+  \<0,+|\hat s_j|0,-\>  {B_x- i B_y \over  B_z+(v-L)^2 \Sigma^+(z_{\xi''})-z_{\xi''} }  \notag \\
& +  \<0,-|\hat s_j|0,-\>  {B_x^2 +B_y^2 \over \left(  B_z+(v-L)^2 \Sigma^+(z_{\xi''})-z_{\xi''} \right) \left(  B_z+(v-L)^2 \Sigma^+(z_{\xi})-z_{\xi} \right) } \Bigg\} \;, \\
\<\phi_{\xi'}|\hat s_i|\phi_{\xi''}\>&={1\over {\cal N}_{\xi'}^{* 1/2} }{1\over {\cal N}_{\xi''}^{1/2} }  \Bigg\{ \<0,+|\hat s_i|0,+\>  \notag  \\
&+  \<0,+|\hat s_i|0,-\>  {B_x+ i B_y \over  B_z+(v-L)^2 \Sigma^+(z_{\xi''})-z_{\xi''} }   \notag \\
&+  \<0,+|\hat s_i|0,-\>  {B_x- i B_y \over  B_z+(v-L)^2 \Sigma^-(z^*_{\xi'})-z^*_{\xi'} }  \notag \\
& +  \<0,-|\hat s_i|0,-\>  {B_x^2 +B_y^2 \over \left(   B_z+(v-L)^2 \Sigma^+(z_{\xi''})-z_{\xi''} \right)  \left(  B_z+(v-L)^2 \Sigma^-(z^*_{\xi'})-z^*_{\xi'} \right) } \Bigg\} \;, \\
\<0,B_+| \tilde\phi_{\xi'} \> &={1\over {\cal N}_{\xi'}^{* 1/2} }   \left\{ \<0,B_+ |0,+\>+\<0,B_+|0,-\> {B_x+ i B_y \over  B_z+(v-L)^2 \Sigma^-(z^*_{\xi'})-z^*_{\xi'}} \right\} \;.
\end{align}
\end{subequations}
%--

\end{widetext}

%####################################
\section{Analytic continuation for the Liouville pathways} \label{AppSec:LiouvillePath}

In this section, we note the direction of the analytic continuation of the Liouville pathway in the double-sided Feynman diagram, where we choose the pathway which is decaying in the future.
Now we consider the transition probability from  state $|a\>$ to $|b\>$ as
%--
\begin{align}
P_{ba}(t)=\left| \<b|e^{-i \hat H t}|a\>\right|^2 \;.
\end{align}
%--
In terms of the Liouville pathway representation, we write this as
%--
\begin{align}\label{AppEq:Pbat}
P_{ba}(t)=\<\!\<b;b|e^{-i {\cal L} t}|a;a\>\!\> \;.
\end{align}
%--
We choose the analytic continuation so that the probability decays for the future.
Therefore, we take the Liouville basis given by\cite{Ordonez01PRA}
%----
\begin{align}\label{AppEq:Lbasis}
|\phi_\xi;\phi_{\xi'}\>\!\>\equiv |\phi_\xi\>\<\phi_{\xi'}| \;, \; \<\!\<\tilde \phi_\xi;\tilde\phi_{\xi'}|\equiv |\tilde\phi_{\xi'}\>\<\tilde\phi_{\xi}| \;,
\end{align}
%--
where $\<\phi_{\xi'}|$ and $\<\tilde\phi_{\xi}|$ are hermitian conjugates of $|\phi_{\xi'}\>$ and $|\tilde\phi_{\xi}\>$, respectively.
While $|\phi_\xi\>$ and $\<\tilde\phi_\xi|$ are the right- and the left-complex eigenstates of the Hamiltonian, the Liouville basis are the complex eigenstates of the Liouvillian:
%--
\begin{align}
{\cal L}|\phi_\xi;\phi_{\xi'}\>\!\>=\Delta_{\xi,\xi'}|\phi_\xi;\phi_{\xi'}\>\!\> \;,\;  \<\!\<\tilde \phi_\xi;\tilde\phi_{\xi'}|{\cal L}=\Delta_{\xi,\xi'} \<\!\<\tilde \phi_\xi;\tilde\phi_{\xi'}| \;,
\end{align}
%--
where
%--
\begin{align}
\Delta_{\xi,\xi'}=z_\xi-z_{\xi'}^* \;.
\end{align}
%--
This basis satisfy the bi-completeness and the bi-orthonormality in the Liouville space as
%--
\begin{align}\label{AppEq:LiouvillComp}
1=\sum_{\xi,\xi'} |\phi_\xi;\phi_{\xi'}\>\!\> \<\!\<\tilde \phi_\xi;\tilde\phi_{\xi'}| \;,\;  
\<\!\<\tilde \phi_\xi;\tilde\phi_{\xi'}|\phi_{\xi''};\phi_{\xi'''}\>\!\> =\delta_{\xi,\xi''}\delta_{\xi',\xi'''} \;.
\end{align}
%--
Substituting (\ref{AppEq:LiouvillComp}) into (\ref{AppEq:Pbat}) we find
%--
\begin{align}
P_{ba}(t)=\sum_{\xi,\xi'}e^{-i (z_\xi-z_{\xi'}^*)t} \<b|\phi_\xi\>\<\phi_{\xi'}|b\> \<\tilde\phi_\xi|a\>\<a|\tilde\phi_{\xi'}\> \;,
\end{align}
%--
which decays with the rate ${\rm Im}(z_\xi-z^*_{\xi'})$ from the contributions of the  resonance and anti-resonance eigenstates, as required.

%#########################

\end{document}